\documentclass[preprint,preprintnumbers,amsmath,amssymb]{revtex4}
\usepackage{amssymb,latexsym,amsmath}

\usepackage[activeacute,english]{babel}
\usepackage{fancyhdr}
\usepackage{textcomp}
\usepackage{amsfonts}
\usepackage{graphicx}
\usepackage{graphics}
\usepackage{subfigure}
\usepackage{color}
\usepackage{array}

\usepackage[active]{srcltx}

\newcommand{\al}{\alpha}
\newcommand{\pa}{\partial}

\newcommand{\la}{\lambda}

\newcommand{\om}{\omega}

\newcommand{\De}{\Delta}

\newcommand{\rar}{\rightarrow}

\begin{document}

\title{Two charges on plane in a magnetic field I. ``Quasi-equal" charges and neutral quantum system at rest cases}

\author{M.A.~Escobar-Ruiz}
\email{mauricio.escobar@nucleares.unam.mx}
\author{A.V.~Turbiner}
\email{turbiner@nucleares.unam.mx}
\affiliation{Instituto de Ciencias Nucleares, Universidad Nacional
Aut\'onoma de M\'exico, Apartado Postal 70-543, 04510 M\'exico,
D.F., Mexico}

\date{October 8, 2013}

\begin{abstract}
Low-lying bound states for the problem of two Coulomb charges of finite masses on a plane subject to a constant magnetic field $B$ perpendicular to the plane are considered. Major emphasis is given to two systems: two charges with the equal charge-to-mass ratio (quasi-equal charges) and neutral systems with concrete results for the Hydrogen atom and two electrons (quantum dot).

It is shown that for these two cases, but when a neutral system is at rest (the center-of-mass momentum is zero), some outstanding properties occur: in double polar coordinates in CMS $(R, \phi)$ and relative $(\rho, \varphi)$ coordinate systems (i) the eigenfunctions are factorizable, all factors except for $\rho$-dependent are found analytically, they have definite relative angular momentum, (ii) dynamics in $\rho$-direction is the same for both systems being described by a funnel-type potential; (iii) at some discrete values of dimensionless magnetic fields $b \leq 1$ the system becomes {\it quasi-exactly-solvable} and a finite number of eigenfunctions in $\rho$ are polynomials. The variational method is employed. Trial functions are based on
combining for the phase of a wavefunction (a) the WKB expansion at large distances, (b) the perturbation theory at small distances (c) with a form of the known analytically (quasi-exactly-solvable) eigenfunctions. Such a form of trial function appears as a compact uniform approximation for lowest eigenfunctions.
For the lowest states with relative magnetic quantum numbers $s=0,1,2$ this approximation gives not less than $7$ s.d., $8$ s.d., $9$ s.d., respectively, for the total energy $E(B)$ for magnetic fields
$0.049\, \text{a.u.} < B < 2000\, \text{a.u.} $ (Hydrogen atom) and $0.025\, \text{a.u.}\eqslantless B \eqslantless 1000\, \text{a.u.} $ (two electrons). The evolution of nodes of excited states with the magnetic field change is indicated. In the framework of convergent perturbation theory the corrections to
proposed approximations are evaluated.
\end{abstract}

\pacs{31.15.Pf,31.10.+z,32.60.+i,97.10.Ld}

\maketitle

\begin{center}
\section*{Introduction}
\end{center}

It is well known that two dimensional, planar quantum systems exhibit many interesting properties both from the point of view of theory and potential applications.
We consider two planar two-body Coulomb systems, $(e_1, m_1)$ and $(e_2, m_2)$, (a) particles with the same charge-to-mass ratio, $\frac{e_1}{m_1} = \frac{e_2}{m_2}$, and (b) particles with opposite charge and arbitrary masses (neutral system), both subject to a perpendicular constant magnetic. These two particular systems play a very important role in different physical sciences. It is worth mentioning that a progress in growing of artificial atoms or quantum dots in semiconductor
heterostructures with a large but finite number of electrons opens a new perspective in fabrication of nanoelectronic devices. The presence of a magnetic field reveals a new physics phenomena, which are absent in the standard atomic-molecular physics, in particular, the existence
of bound states of two electrons is a remarkable example of this (see e.g. \cite{Hall}).
The Hydrogen atom in a strong magnetic field has relevance coming from astrophysics, where
spectra in strong fields have been known since at least 1970, e.g. in the strongly magnetic white dwarf stars. Excitons and shallow impurities in semiconductors reveal hydrogen-like spectra.
Consequently, several analytical, variational and numerical studies have been developed in the literature \cite{GD:1967}-\cite{Burkova} in both 2D and 3D cases. Unlike the present paper, the previous studies of Hydrogen atom in 2D the present authors are familiar with, do not take into account the finite mass effects. There
are explicit indications that in the case of finite masses new phenomena occurs in both classical and quantum
problems (see e.g. \cite{Turbiner} - \cite{Taut2} and references therein).

We are going to use the variational method with physically adequate trial functions \cite{Turbiner:1979-84}. To choose such trial functions we employ a simple idea to combine a WKB expansion at large distances with perturbation theory at small distances near the extremum of the potential into an interpolation similar to one
already successfully used for $1D$ anharmonic oscillator \cite{Turbiner:2005-10}. These interpolations
turned out to be quite accurate uniform approximations of the exact eigenfunctions.
A particular goal of this paper is to present such an approximation for several lowest states of the Hydrogen atom and two electrons.

\section{Generalities}

The Hamiltonian, which describes a two-body system in a constant and uniform magnetic field ${\bf B} = B\ \bf \hat z$ perpendicular to the plane, is of the form
(assuming $\hslash=\frac{1}{4\,\pi\, \epsilon_0}=1$),
\begin{equation}
\begin{aligned}
{ {\hat H}} = &  \frac{{({\mathbf {\hat p}_1}-\frac{e_1}{c}\,{\mathbf A_1})}^2}{2\,m_1} + \frac{{({\mathbf {\hat p}_2}-\frac{e_2}{c}\,{\mathbf A_2})}^2}{2\,m_2}
+ \frac{e_1\,e_2}{\mid {\boldsymbol \rho}_1 - {\boldsymbol \rho}_2 \mid} \,,\qquad \boldsymbol{\rho}_{1,2} \in \Re^2\ ,
\label{Hcar}
\end{aligned}
\end{equation}
where ${\mathbf {\hat p}}_{1,2}=-i\,\nabla_{1,2}$ is the momentum and ${\boldsymbol \rho}_{1,2}$ is the position of the first (second) particle. We assume the symmetric gauge $\bf A_{1,2}=\frac{1}{2}\,\bf B\times {\boldsymbol \rho}_{1,2}$ is chosen. It is easy to check that the Pseudomomentum,
\begin{equation}
   {\bf {\hat K}}\ =\ {\bf {\hat p}_1}  +  \frac{e_1}{c}\,{\bf A_{1} + {\hat p}_2} + \frac{e_2}{c}\,{\bf A_{2}}\,.
\label{pseudo}
\end{equation}
is a vector integral of motion in the plane, where the dynamics is developed,
\[
      [ {\hat H}\ ,\ {\bf {\hat K}}]\ =\ 0\ ,
\]
as well as the total angular momentum
\begin{equation}
\boldsymbol  {\hat L} = {\boldsymbol \rho}_1 \times {\mathbf {\hat p}}_1+ {\boldsymbol \rho}_2\times {\mathbf {\hat p}}_2\ ,
\label{Lz}
\end{equation}
$[\, \boldsymbol  {\hat L}, \,  {\hat H} \,]=0$. The vector $\boldsymbol  {\hat L}$ is perpendicular to the plane.
In general, the problem is not completely integrable: the number of mutually commuting integrals (including the Hamiltonian) is less than four, the dimension of the configuration space.

\hskip 1cm
It is convenient to introduce center-of-mass (CMS) and relative coordinates
\begin{equation}
\begin{aligned}
&\mathbf R = \mu_1\, {\boldsymbol \rho}_1 + \mu_2\,{\boldsymbol \rho}_2 \,,
\quad  {\boldsymbol \rho}= {\boldsymbol \rho}_1 - {\boldsymbol \rho}_2\,,
\\ & \mathbf {\hat P} = {\mathbf {\hat p}}_1 + {\mathbf {\hat p}}_2 \,,
\qquad \quad \, \, {\mathbf {\hat p}} = \mu_2\,{\mathbf {\hat p}}_1 -  \mu_1\,{\mathbf {\hat p}}_2\ ,
\end{aligned}
\label{CMvar}
\end{equation}
where $\mu_i=\frac{m_i}{M}$ is reduced mass of the $i$th particle and $M = m_1 + m_2$ the total mass of the system. In these coordinates
\begin{equation}
\mathbf {\hat K}  = \mathbf {\hat P} + \frac{q}{c}\,\mathbf A_{\mathbf R} + \frac{e_c}{c}\,\mathbf A_{{\boldsymbol \rho}}  \ ,
\label{pseudoR}
\end{equation}
\begin{equation}
\boldsymbol {\hat L}\ =\ ({\mathbf R} \times {\bf {\hat P}}) + ({\boldsymbol \rho}\times \bf {\hat p})
\equiv \mathbf {\cal{\hat L}} + \boldsymbol {\hat \ell} \ ,
\label{LzR}
\end{equation}
(cf. (\ref{pseudo}), (\ref{Lz})), where
\[
q = e_1 + e_2\ ,
\]
is the total charge and the \emph{coupling} charge
\[
e_c \ =\ (e_1\,\mu_2-e_2\,\mu_1)\ = \  m_r\,\bigg( \frac{e_1}{m_1} - \frac{e_2}{m_2} \bigg)\ ,
\]
where $m_r$ is the reduced mass of the system, $m_r=\frac{m_1 m_2}{M}$.

\hskip 1cm
The integrals ${\bf {\hat K}} = ({\hat K}_x, {\hat K}_y),\
{\boldsymbol {\hat L}} = {\hat L}\ \bf {n_z}$ obey the commutation relations
\begin{equation}
\begin{aligned}
&[ {\hat K}_x,\,{\hat K}_y ] = -\frac{q\,B}{c}\,,
\\ & [ {\hat L} ,\,{\hat K}_x ] = {\hat K}_y\,,
\\ & [ {\hat L},\,{\hat K}_y ] = -{\hat K}_x\,,
\label{AlgebraInt}
\end{aligned}
\end{equation}
where $\bf {n_z}$ is the unit normal vector to the plane.
Hence, they span a noncommutative algebra with the Casimir operator ${\cal {\hat C}}$,
\begin{equation}
{\cal {\hat C}}\ =\ {\hat K}_x^2+{\hat K}_y^2-\frac{2\,q\,B}{c}{\hat L} \ .
\label{Casimir}
\end{equation}

\hskip 1cm
It is convenient to unitary-transform the canonical momenta
\[
  U^{-1}\,{\bf {\hat P}}\, U \ = \ {\bf {\hat P}} +  \frac{e_c}{c}\,{\bf A}_{\boldsymbol \rho}  \quad , \quad
U^{-1}\,{\bf {\hat p}}\, U \ = \ {\bf {\hat p}} - \frac{e_c}{c}\,{\bf A}_{\bf R}
\ ,
\]
with
\begin{equation}
\label{U}
 {U} = e^{-i\,\frac{e_c}{c}\,\bf A_{\boldsymbol \rho}\cdot \bf R} \,.
\end{equation}
The unitary transformed Pseudomomentum (\ref{pseudoR}) reads
\begin{equation}
  {\mathbf K^{\prime}}\ =\ U^{-1}\,{\bf {\hat K}} \, U \ = \  {\bf {\hat P}} + \frac{q}{c}\,\bf A_{\bf R}   \,,
\label{KTrans}
\end{equation}
and looks like as pseudomomentum of the whole, composite system of charge $q$, see (\ref{pseudo}).
The unitary transformed Hamiltonian (\ref{Hcar}) takes the form
\begin{equation}
  {\cal {\hat H}}^{\prime} \ =\  U^{-1}\,{\cal {\hat H}}\, U \ = \ \frac{ {( \mathbf {\hat P}-\frac{q}{c}\,\mathbf A_{\bf R}-2\,\frac{e_c}{c}\,\mathbf A_{\boldsymbol \rho} )}^2}{2\,M}
  +\frac{{({\mathbf {\hat p}}-\frac{q_\text{w}}{c}\,{\mathbf A_{\boldsymbol \rho}})}^2}{2\,m_{r}} +\frac{e_1\,e_2}{\rho}\ ,
\label{H}
\end{equation}
here $q_{\rm{w}} \equiv e_1\,\mu_2^2+e_2\,\mu_1^2$  is an effective charge (weighted total charge).
It is evident, $[\, \mathbf {\hat K}^{\prime}, \, {\cal {\hat H}}^{\prime} \,]=0$\ . The eigenfunctions of ${\cal {\hat H}}$ and ${\cal {\hat H}}^{\prime}$ are related through a phase rotation
\begin{equation}
   \Psi^{\prime}\ =\ \Psi\ e^{ i\,\frac{e_c}{c}\,\mathbf A_{\boldsymbol \rho}\cdot \mathbf R}\ .
\label{psiprime}
\end{equation}

Below we are going to study the spectra of the Hamiltonian (\ref{H}) for two particular systems,

\bigskip

(i)  $e_c=0$, where separation of c.m.s. variables occurs \cite{Kohn:1961},

\vspace{0.2cm}

(ii) $q=0$, for which components of the Pseudomomentum $\mathbf {\hat K}$ become commutative (see (\ref{AlgebraInt})).

\hskip .8cm
It is worth noting that for these two cases the problem becomes {\it superintegrable} (the number of integrals is larger than the dimension of the configuration space) and {\it quasi-exactly-solvable} (see \cite{Turbiner:1988Z}-\cite{Turbiner:1988}) for some discrete values of magnetic field \cite{Turbiner}, \cite{ET-q}. For these values of magnetic field $B$, some eigenfunctions can be found analytically \cite{Taut1}, \cite{Taut2}, \cite{ET-q}.

\newpage

\section{Quasi-equal charges - charges of equal Larmor frequency $(e_c=0)$}

\hskip 1cm
This case appears for charges of the same sign and equal cyclotron frequency, $\frac{e_1}{m_1}=\frac{e_2}{m_2}$. The Hamiltonian $(\ref{H})$ becomes
\begin{equation}
\begin{aligned}
{\cal {\hat H}^{\prime}}& \  =\ {\cal {\hat H}} =
{\cal {\hat H}}_{R}(\bf {\hat P},\bf R)\ +\
{\cal {\hat H}}_{\rho}(\bf { \hat p},\boldsymbol \rho)
\\& \equiv \ \frac{({\bf {\hat P}}-\frac{q}{c}\,{\bf A}_{\bf R})^2}{2\,M}\ +\
\frac{({\bf {\hat p}} - \frac{e_1\,m_2}{M\,c} {\bf A}_{\boldsymbol \rho} )^2}{2\,m_r}\ +\
\frac{m_2}{m_1}\,\frac{e_1^2}{\rho}\ ,
\end{aligned}
\label{Hec}
\end{equation}
where CMS variables are separated\,. Here ${\cal {\hat H}}_{R}(\bf {\hat P},\bf R)$ and ${\cal {\hat H}}_{\rho}(\bf { \hat p},\boldsymbol \rho)$ describe CMS and relative motion of two-body composite system, respectively, like it appears for field-free case.
It can be easily shown that four operators
\begin{equation}
{\cal {\hat H}}_{R}\,,\,{\cal {\hat H}}_{\rho}\,,\,{\hat {L}}_z\,,\, {\hat {l}}_z\ ,
\label{CSO}
\end{equation}
(see (\ref{LzR}))
are mutually commuting operators spanning a commutative algebra. Hence, at $e_c=0$ the system is completely integrable. Any state is characterized by four quantum numbers.

\hskip 1cm
Due to decoupling of CMS and relative motion in (\ref{Hec}) the eigenfunctions are factorized
\begin{equation}
{\cal {\hat H}} \,\Psi=(E_R+E_\rho) \, \Psi \,,\qquad \Psi=\chi(\mathbf R) \, \psi(\boldsymbol \rho)\,,\qquad \Psi \in L^2(\Re^4)\ .
\label{Psi1}
\end{equation}
The eigenfunctions of the CMS motion are same than those of a one-particle problem (the Landau problem), with charge $q$ and mass $M$, in a constant magnetic field, i.e. it is an exactly solvable problem. In CMS polar coordinates, ${\bf R}=(R,\,\theta)$, we write eigenfunctions (and spectra) in the following form, assuming $e_1>0$,
\begin{equation}
\begin{aligned}
& \chi = R^{| S |}\,{\rm e}^{i\,S\,\theta}\,{\rm e}^{-\frac{M\,\om_c\,R^2}{4}}\,
L_N^{(| S |)}\bigg(\frac{2\,R}{M\,\om_c}\bigg)\  ,
\\ & E_R=\frac{\om_c}{2}(2\,N+1+|S|-S)\,, \qquad \om_c=\frac{e_1\,B}{m_1\,c} \qquad \ ,
\end{aligned}
\label{Psiec}
\end{equation}
where $L_N^{(|S|)}$ is associated Laguerre polynomial with index $|S|$, $N=0,\,1,\,2...$ is the principal quantum number and $S=0, \pm 1,\,\pm 2,...$ is the CMS magnetic quantum number.
Notice that the CMS motion cyclotron frequency is equal to one of the individual charge.
Eventually, the spectra of  Pseudomomentum (\ref{KTrans}),
\[
    K^2\ =\ \frac{q\, B}{c}\, (2\,N + 1 + |S| + S)\ .
\]

\hskip 1cm
Let us proceed to study the relative Hamiltonian, ${\cal {\hat H}}_\rho\, \psi= E_\rho \, \psi$, see (\ref{Hec}). Their eigenfunctions admit a factorization in relative polar coordinates $\boldsymbol \rho=(\rho,\,\varphi)$,

\vspace{0.2cm}

\begin{equation}
\begin{aligned}
&\psi(\boldsymbol \rho) = {\rm e}^{i \,s\, \varphi}\ \text{e}^{-\frac{e_1\,m_{2}\,\,B}{4\,(m_{1}+m_{2})\,c}\,\rho^2}\ \rho^{|s|}\ p(\rho)\,,
\label{psia}
\end{aligned}
\end{equation}

\vspace{0.2cm}

where $s=0,\,\pm 1,\,\pm 2, \ldots$ is the magnetic quantum number, where the function $p(\rho)$ obeys

\vspace{0.3cm}

\begin{equation}
\begin{aligned}
 &   - \pa_\rho^2\,p  + \bigg(\om_c\,m_r\,\rho - \frac{1+2\,|s|}{\rho}\bigg)\,\pa_\rho\,p \ +\
 2 \bigg(\ \frac{\,m_r\,m_2\,e_1^2}{m_1\,\rho}\ -\ \hat E_\rho m_r\bigg)\,p  \ = \ 0 \ ,
\label{eqa}
\end{aligned}
\end{equation}
where the reference point for the energy is changed
\[
 \hat E_\rho = E_\rho - \frac{\om_c}{2}\,(1+ |s|-s)\ .
\]
The boundary conditions for (\ref{eqa}) are chosen in such a way that
\[
   p(\rho) \rar \mbox{const}\quad \mbox{at}\quad \rho \rar 0\quad ,\quad p(\rho) \text{e}^{-\frac{e_1\,m_{2}\,\,B}{4\,(m_{1}+m_{2})\,c}\,\rho^2}\ \rho^{|s|} \rar 0\quad \mbox{at}\quad \rho \rar \infty\ .
\]
It is clear that $\hat E_\rho =\hat E_\rho(e_1,\,m_1,\,m_2,\,B,\,s)\ $.
This equation can be transformed into $1D$ Schr\"odinger equation for the funnel-type potential, $\frac{A}{\rho} + B \rho + C \rho^2$. About this potential it is known that for a certain combination of parameters $A, B, C$ analytic eigenfunctions occur in a form of polynomial multiplied by some factor \cite{Turbiner:1988Z}, \cite{Turbiner:1988}, \cite{Turbiner}. Hence, for specific values of a magnetic field $B_n$, (\ref{eqa}) possesses polynomial solutions $P_n(\rho)$, $n=1,2,3,..$\, \cite{ET-q}. In particular, for the case of two identical particles, ($e_1=e_2\equiv e\,;\,m_1=m_2\equiv m$), the largest magnetic field $B_{max}$ for which the problem (\ref{eqa}) admits an analytical solution is
\[
     B_{max}\ =\ B_1\ =\ B_0 \equiv \,2\,m^2\,e^3\,c \ ,
\]
(see e.g. \cite{Taut1}), where $B_0$ is the characteristic magnetic field which defines the magnetic field unit. The corresponding eigenfunction is equal to
\begin{equation}
P_1 \ = \ 1 + \sqrt{\frac{e\, B_1}{2\,c}}\,\rho\,,
\label{psi1a}
\end{equation}
which corresponds to the ground state (at relative magnetic quantum number $s=0$) with the energy
$E_\rho =  \frac{e\,B_1}{c\,m}$.

\hskip 1cm
The appearance of quasi-exact-solvability in the relative motion can be interpreted as appearance of  the extra (particular) integral of motion for a certain values of a magnetic field. Let us
denote an analytic eigenfunction as $\psi^{qes}({\boldsymbol \rho})$. It implies that
\[
{\cal {\hat H}}_\rho \, \psi^{qes}({\boldsymbol \rho})  \ =\
\frac{e_1\,B}{2\,m_1\,c}\,(n+1+|s|-s)  \, \psi^{qes}({\boldsymbol \rho})\ .
\]
One can construct the operator
\begin{equation}
{I}_{n,|s|}\ =\ \tau \, \prod_{j=0}^n(\rho\,\pa_\rho-j)\,\tau^{-1}\ =\ \prod_{j=0}^n(\rho\,{\cal D}_\rho-j)\ ,
\label{Inec}
\end{equation}
where the gauge factor $\tau = \rho^{|s|}\ \text{e}^{-\frac{e_1\,m_{2}\,\,B}{4\,(m_{1}+m_{2})\,c}\,\rho^2}$ and the covariant derivative
\[
{\cal D}_\rho\ =\  \pa_\rho  + \frac{m_r\,\om_c\,\rho}{2} - \frac{|s|}{\rho} \ .
\]
It annihilates $\psi^{qes}$,
\[
    {I}_{n,|s|} \psi^{qes} \ =\ 0\ .
\]
Thus, $\psi^{qes}$ is zero mode of ${I}_{n,|s|}$.
It immediately implies that
\begin{equation}
[{\cal {\hat H}}_\rho\,,\, {I}_{n,|s|} ]\, \psi^{qes}({\boldsymbol \rho}) =0\,,
\label{inteec}
\end{equation}
It is evident that the operator ${I}_{n,|s|}$ has $(n+1)$ zero modes and for each of them the equation (\ref{inteec}) holds. Thus, the commutator $[{\cal {\hat H}}_\rho\,,\, {I}_{n,|s|}]$ vanishes on the space of zero modes of ${I}_{n,|s|}$. Therefore, ${I}_{n,|s|}$ is
a particular integral (for discussion see \cite{Turbiner:2013}).

\textbf{Scaling relations.}

Let us consider two $(e_c=0)$ systems, $(e, m_1, m_2)$ and $(\tilde e, \tilde m_1, \tilde m_2)$.
Making scale transformation in (\ref{eqa}), $\rho \rar a \rho$, one can arrive at a certain relation between eigenstates. If
\[
    a\ =\ \frac{\tilde e^2}{e^2}\ \frac{m_1}{\tilde m_1}\ \frac{\tilde m_2}{m_2}\
    \frac{\tilde m_r}{m_r}\ ,
\]
and
\[
    \tilde B\ =\ \frac{\tilde e^3}{e^3}\ \frac{m_1}{\tilde m_1}\ \frac{\tilde m_2^2}{m_2^2}\ \frac{\tilde m_r}{m_r}\ B\ ,
\]
then the following scaling relations emerge
\[
  \psi(\tilde e,\,\tilde m_1,\,\tilde m_2,\,\tilde B,\, s; a\rho) \ =\ \psi(e,\,m_1,\,m_2,\,B,\,s; \rho)\ ,
\]
\begin{equation}
\frac{\tilde m_1^2}{\tilde e^4\,\tilde m_r\,\tilde m_2^2}\,
{\hat E}_\rho(\tilde e,\,\tilde m_1,\,\tilde m_2,\,\tilde B,\, s)
\ =\
\frac{m_1^2}{e^4\, m_r\,m_2^2} \,
{\hat E}_\rho(e,\,m_1,\,m_2,\,B,\,s)
\ .
\label{scalec}
\end{equation}
It is worth noting how the scaling relations look like for a particular case of equal charges but proportional masses, $(e, m_1, m_2)$ and $(e, b\, m_1, b\, m_2)$,
\[
  \psi(e,\, b m_1,\,b m_2,\,b^2 B,\, s; b \rho) \ =\ \psi(e,\,m_1,\,m_2,\,B,\,s; \rho)\ ,
\]
\begin{equation}
{\hat E}_\rho(e,\,b m_1,\,b m_2,\,b^2 B,\, s)
\ =\
b \,{\hat E}_\rho(e,\,m_1,\,m_2,\,B,\,s) \ .
\label{scalec-p}
\end{equation}

\textbf{Asymptotics.}

Making the analysis of Eq. (\ref{eqa}) we arrive at
\begin{equation}
      p\ = \ 1 +c_1\, \rho + c_2\,\rho^2 + \ldots \ ,\quad \rho \rar 0\ ,
\label{rho+a}
\end{equation}
with
\[
      c_1 \ = \ \frac{2\,e_1^2\,m_2\,m_r}{m_1\,(1+2\,|s|)}\ ,
\]
\[
      c_2 \ = \ \frac{1}{2\,(4\,s^2 + 6\,|s| + 2  )}\bigg[ \frac{4\,e_1^4\,m_2^2\,m_r^2}{m_1^2} + (1+2\,|s|) \bigg( \frac{e_1\,B\,m_r\,(1+|s|-s)}{c\,m_1} - 2\,E_\rho\,m_r  \bigg)  \bigg]\ ,
\]
which is, in fact, the perturbation theory expansion near the minimum of the funnel-type potential at $\rho=0$. From another end, the expansion at large $\rho \rightarrow \infty $ (WKB asymptotics) has the form
\begin{equation}
 p\ = \ \rho^\beta(1 + \frac{C_1}{\rho} + \frac{C_2}{\rho^2}\ +\ \ldots)\ ,
 \quad \beta\ =\ -1 + s - |s| + \frac{2\,c\,E_\rho\,m_1}{B\,e_1} \ ,
\label{rho+b}
\end{equation}
where
\[
      C_1 \ = \ \frac{2\,c\,e_1\,m_2}{B}\ ,
\]
\[
      C_2 \ = \ \frac{c}{2\,B^3\,e_1^3\,m_r}\bigg[-4\,c^2\,E_\rho^2\,m_1^3   + 4\,B\,c\,e_1\,(e_1^4\,m_2^2\,m_r - E_\rho\,m_1^2(s-1)) + B^2\,e_1^2\,m_1\,(2\,s-1)      \bigg]\ .
\]

\hskip 0.8cm

At large $B$ the ground state the ground state energy behaves
\[
E_0\ =\ \frac{e_1\,m_2}{2\,c\,M m_r}\ B \, -\,\al \sqrt{\frac{e_1 m_2\,\pi}{2\,c M}}\ B^{1/2}\, +\, \ldots \ ,
\]
where $\al = e_1 e_2 = \frac{m_2}{m_1} e_1^2$\ \ is the strength of Coulomb interaction.

\subsection{Approximations}


\hskip 1cm
One can make an interpolation for $p$ between perturbation theory (\ref{rho+a}) and the WKB expansion (\ref{rho+b}) keeping in mind a form of the exact solutions of (\ref{eqa}) (see \cite{ET-q}) which must emerge for specific values of magnetic field \cite{Turbiner:2005-10}. The simplest interpolation for lowest states has the following forms:

(i) the ground state

\vspace{0.2cm}

\begin{equation}
 p_0\ =\ {\bigg(\frac{1+a_1\,\rho + a_2\,\rho^2+a_3\,\rho^3}{1+a_4\,\rho}\bigg)}^{\al_0}\ ,
\label{grounda}
\end{equation}

\vspace{0.2cm}

(ii) one-node state,

\vspace{0.2cm}

\begin{equation}
p_{1}\ = {(1 + b_1\,\rho + b_2\,\rho^2)}^{\al_1}\,(1 - b_3\,\rho)\ ,
\label{1THa}
\end{equation}

\vspace{0.2cm}

(iii) two-node states,

\vspace{0.2cm}

\begin{equation}
p_{2}\ = {(1 + d_1\,\rho + d_2\,\rho^2)}^{\al_2}\,(1 - d_3\,\rho + d_4\,\rho^2)\ ,
\label{2THa}
\end{equation}

\vspace{0.2cm}

where $a^\prime s,\,b^\prime s,d^\prime s$ and $\al^\prime s$ are parameters which will be found variationally. Supposedly, they should behave smoothly as a function of a magnetic field. Parameters $\al_0, (\al_1+1), (\al_1+2)$ should be close to $\beta$ (see (\ref{rho+b})).
For a discrete values of magnetic field the function $p(\rho)$ reduces to the corresponding exact solution. An information about the nodes is coded in the factor $(1 - b_3\,\rho)$ for the first excited state (\ref{1THa}) and $(1 - d_3\,\rho + d_4\,\rho^2)$ for the second excited state (\ref{2THa}), respectively.
As usual in variational studies, the orthogonality conditions between states are imposed which effectively reduce the number of free parameters.


\subsection{Results}

\begin{figure}[htp]
\begin{center}
\includegraphics[width=3.5in,angle=-90]{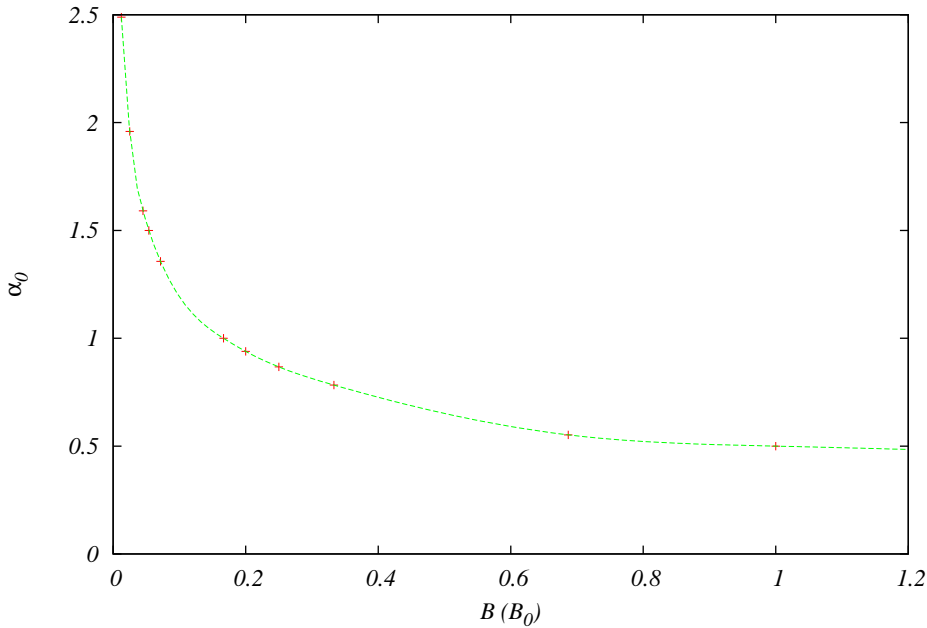}
\caption{Ground state for two-electron case: optimal value of the parameter $\al_0$ in (\ref{grounda}) {\it vs} magnetic field $B$\ .}
\label{ec-alpha0}
\end{center}
\end{figure}

\hskip 1cm
Below we explore the case of two electrons, $e_1=e_2, m_1=m_2$, which is the most important particular case of quasi-equal charges, $e_c=0$. Other cases of $e_c=0$ can be studied through scaling relations (\ref{scalec}) - (\ref{scalec-p}). We limit ourselves to a consideration of several low-lying states $(N,S; n,s)$ with relative quantum number $n=0,1,2$ and relative magnetic quantum number $s=0,1,2$, see Tables \ref{Table1} - \ref{Table5} at arbitrary center-of-mass quantum numbers $N,S$. The first observation is that the very simple, few-parametric, variational trial functions (\ref{grounda}), (\ref{1THa}), (\ref{2THa}) lead to extremely accurate variational energies, see Tables \ref{Table1}, \ref{Table2}, \ref{Table4} as well as position of nodes, see Tables \ref{Table3}, \ref{Table5}. Optimal parameters depend smoothly on a magnetic field $B$, changing slowly with its variation. On Fig.\ref{ec-alpha0} the behavior of the parameter $\al_0$ in (\ref{grounda}) is shown. At magnetic field values $B=\frac{1}{10+\sqrt{73}},\ 1/6,\ 1$  which correspond to the appearance of the exact solutions \cite{ET-P:2013}, as a result of variational calculation $\al_0$ takes values 1/2, 1, 3/2, respectively, where $p_0$ becomes a polynomial. For magnetic fields tending to zero the optimal $\al_0$ tends to infinity. It reflects the non-existence of the bound state of two electrons for vanishing magnetic field. At large $B$ the optimal $\al_0$ monotonically tends to zero. The energies decrease linearly with a magnetic field decrease to zero. Similarly the energies grow linearly with a magnetic field increase.

\hskip 1cm
A natural question to pose is about the accuracy of obtained variational results, in particular, how close a variational trial function is to the exact one. In order to answer the question a convergent perturbation theory is used \cite{Turbiner:1979-84} (see Appendix A) with the variational trial function as zero approximation. It allows us to estimate a deviation of the variational trial function from the exact one.

\begin{figure}[htp]
\centering
\subfigure[]{\includegraphics[width=2.5in,angle=0]{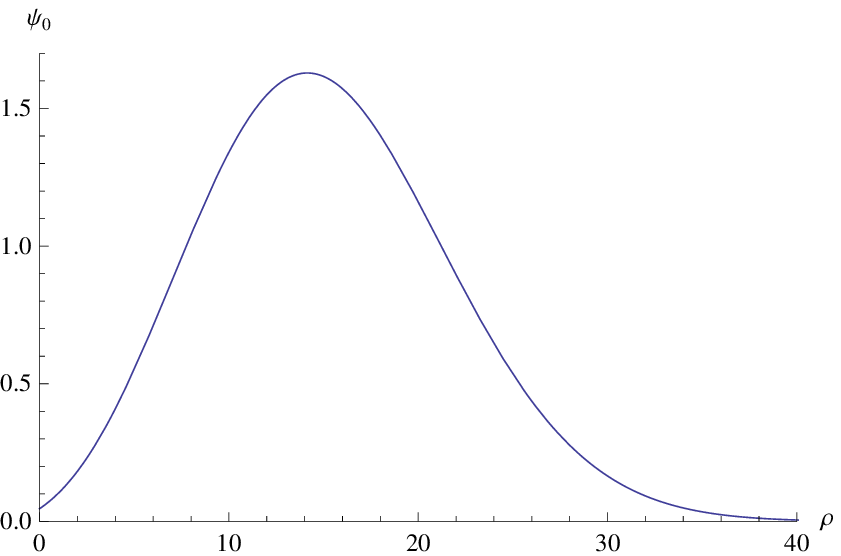}} \qquad \subfigure[]{\includegraphics[width=2.5in,angle=0]{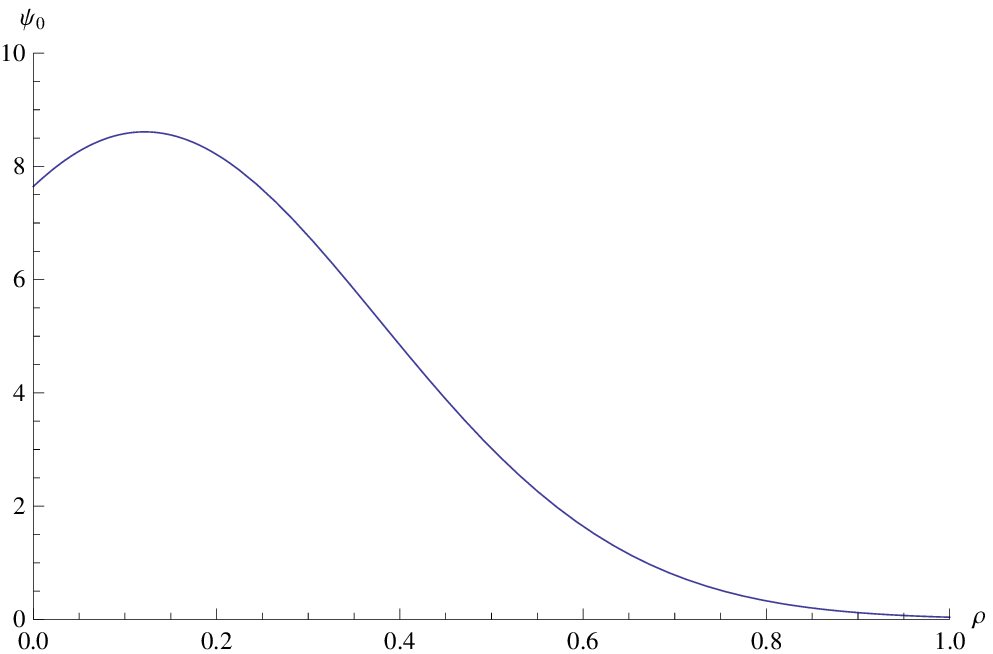} }
\vspace{0.5cm} \\
\subfigure[]{\includegraphics[width=2.5in,angle=0]{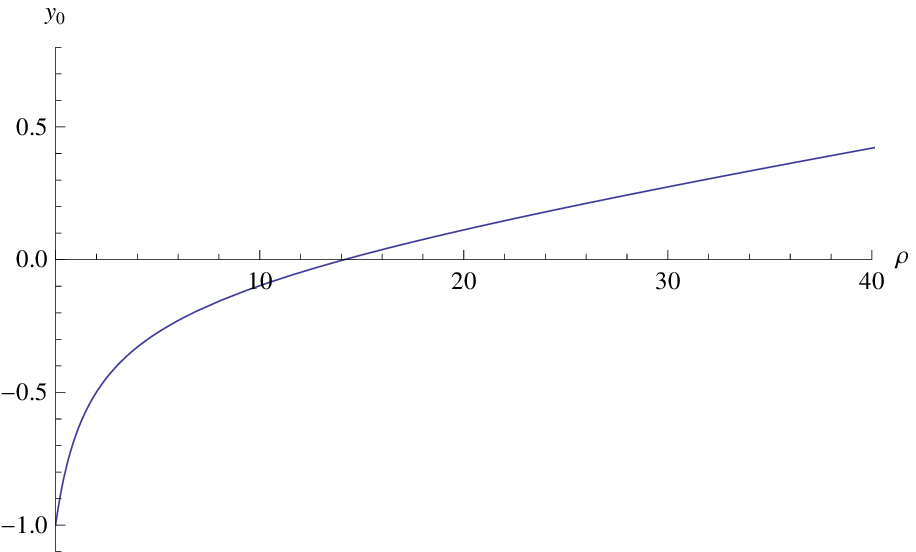}} \qquad \subfigure[]{\includegraphics[width=2.5in,angle=0]{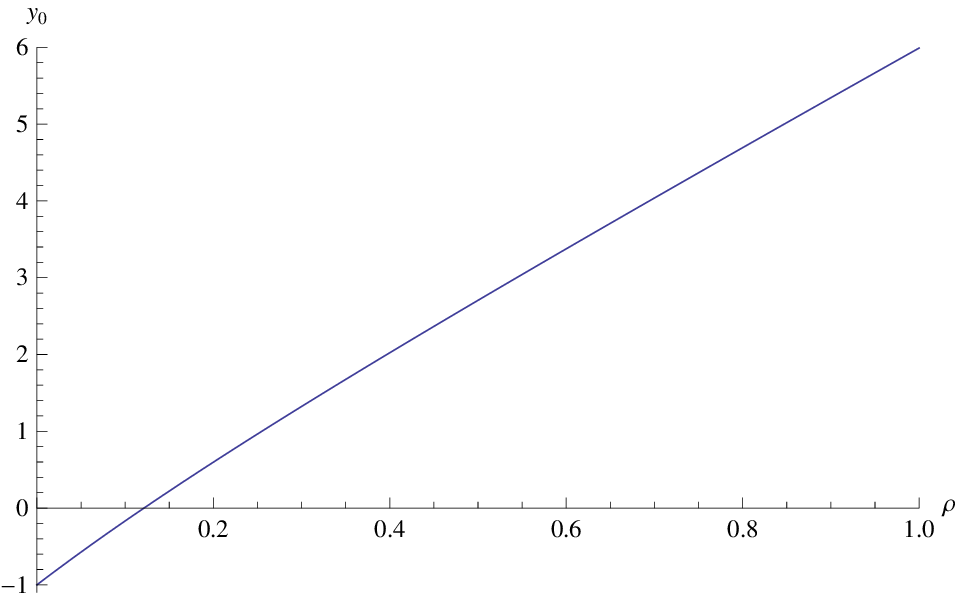} }
\vspace{0.5cm} \\
\subfigure[]{\includegraphics[width=2.6in,angle=0]{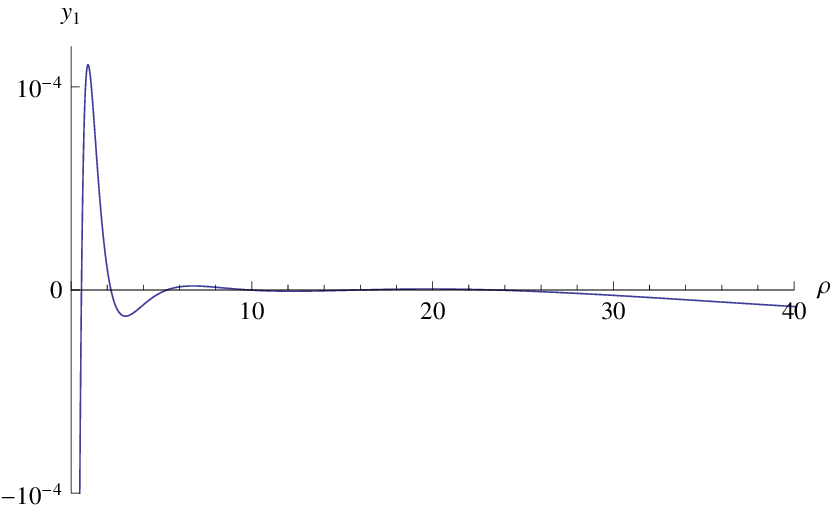}} \qquad \subfigure[]{\includegraphics[width=2.6in,angle=0]{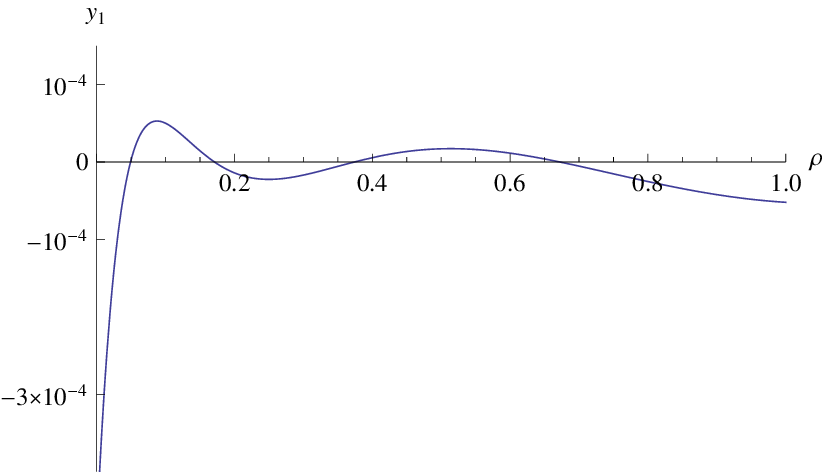} }
\caption{Ground state for two electrons at $s=0$, at $B=\frac{1}{5}B_0$ (left)
and $B=25\,B_0$ (right):
(a)-(b) eigenfunction (\ref{psia}), with prefactor (\ref{grounda}),
taken as zero approximation $\psi \equiv e^{-\phi_0}$,
(c)-(d) $y_0(\rho) = (\phi_0)'$ and (e)-(f) the first correction
$y_1(\rho)$ (see (\ref{yn}))\ .\
$B_0 = 4.701\times 10^9 \,G$. }
\label{y1ec}
\end{figure}

\hskip 1cm
Analysis of the first correction to the eigenfunction allows us to draw a conclusion that the trial functions (\ref{grounda}), (\ref{1THa}), (\ref{2THa}) at optimal values of
parameters are very accurate uniform approximations of the exact eigenfunction. Locally, the approximation provides at least 3-5 significant digits (s.d.)
exactly for any value of the external magnetic field strength(!), (see Fig. \ref{y1ec}). Furthermore, for a domain which gives a dominant contribution to energy integral, $\langle \psi_{trial} {\cal H} \psi_{trial} \rangle$ the number s.d. increases to 9-10.
It is the reason why the variational energy gets known with 8-10 significant digits.

\begin{figure}[htp]
\begin{center}
\includegraphics[width=3.5in,angle=-90]{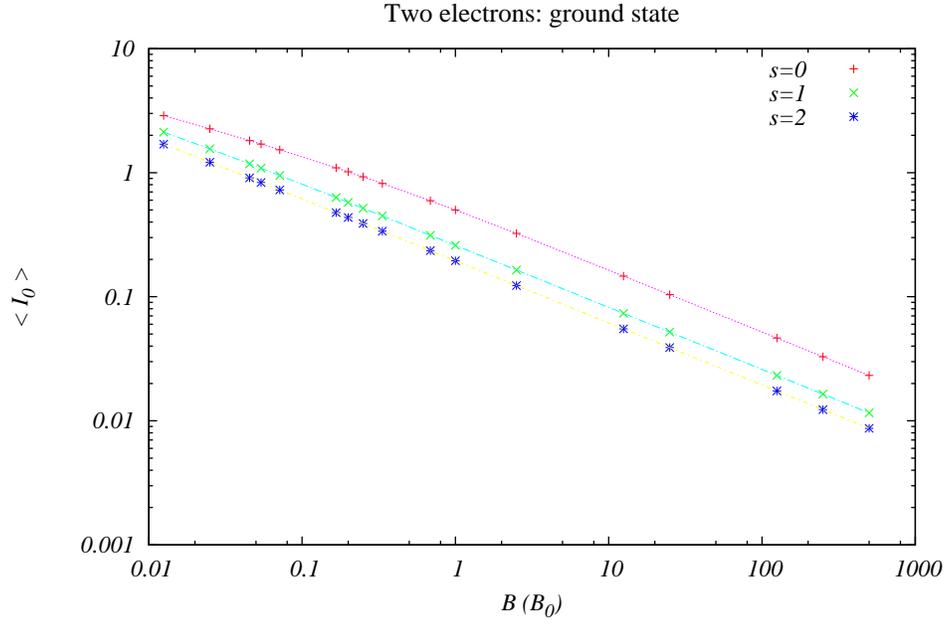}
\caption{Expectation value $\langle {I}_{0,|s|} \rangle$, see (\ref{Inec}) at $s=0,1,2$
{\it vs} magnetic field $B$, $B_0 = 4.701 \times 10^9 G\ .$
\label{Fig.1}}
\end{center}
\end{figure}

\begin{figure}[htp]
\begin{center}
\includegraphics[width=3.5in,angle=-90]{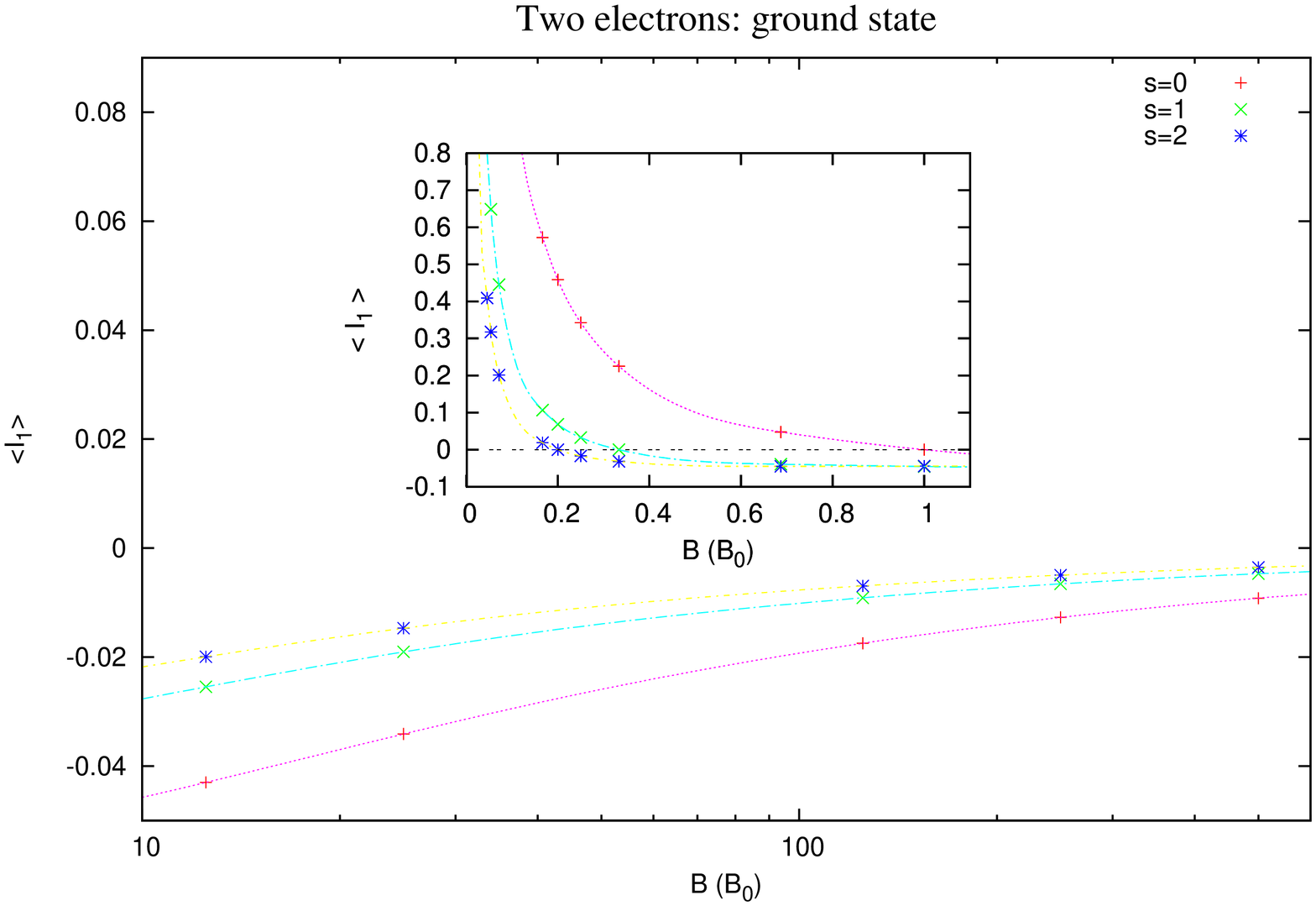}
\caption{Expectation value $\langle {I}_{1,|s|} \rangle$, see (\ref{Inec}) at $s=0,1,2$
vs magnetic field $B$. For $B=\frac{1}{1+2\,|s|}\ B_0 $, the analytic solutions
in (\ref{eqa}) occur at $n=1$ and $\langle {I}_{1,|s|} \rangle$ vanishes.
$B_0 = 4.701  \times 10^9 G\ .$
\label{Fig.2}}
\end{center}
\end{figure}

On Figs. \ref{Fig.1} - \ref{Fig.2} the expectation values for the particular integrals ${I}_{0,|s|}$ and ${I}_{1,|s|}$ at $s=0,1,2$, see (\ref{Inec}), are shown. Note that at $B_s=\frac{1}{1+2\,|s|}\ B_0$ the expectation value $\langle {I}_{1,|s|} \rangle$ vanishes. It corresponds to appearance of the analytic solution in (\ref{eqa}) at $n=1$.

\clearpage

\section{neutral system ($q=0$) at rest $(\bf P=0)$}

\hskip 1cm
Let us take a neutral system, $(e, m_1), (-e, m_2)$, assuming $e>0$.  In this case the unitary transformed Hamiltonian $(\ref{H})$ takes the form
\begin{equation}
 {\cal H}^{\prime}\ =\ \frac{ {( \mathbf {\hat P}-2\,e\,\mathbf A_{\boldsymbol \rho} )}^2}{2\,M} +
 \frac{{({\mathbf {\hat p}}-e\,(\mu_2-\mu_1)\,{\mathbf A_{\boldsymbol \rho}})}^2}{2\,m_{r}}  -\frac{e^2}{\rho} \ .
\label{HK}
\end{equation}

CMS in (\ref{HK}) is not separated out. Unlike the case $e_c=0$, see Section II, the Hamiltonian is not completely integrable. The unitary transformed total Pseudomomentum (\ref{KTrans}), which commutes with ${\cal H}^{\prime}$, coincides with CMS momentum,
\[
\mathbf {\hat K}^{\prime} = \mathbf {\hat P}\ .
\]
It is easy to check that the eigenfunction of $\mathbf {\hat P}$ has the form
\begin{equation}
 \Psi^{\prime}_{{}_{\bf P}}(\mathbf R\,, \boldsymbol \rho)\ = \ \text{e}^{i\,\bf P\cdot \bf R}\,\psi_{{}_{\bf P}}(\boldsymbol \rho)    \ ,
\label{psik}
\end{equation}
where $\bf P$ is the eigenvalue and $\psi_{{}_{\bf P}}(\boldsymbol \rho)$ depends on the relative coordinate $\boldsymbol \rho$.

\hskip 1cm
Studying the classical neutral system we found special, superintegrable, closed trajectories at a vanishing Pseudomomentum, $\bf P=0$, see (\cite{ET:2013}). It implies that besides the global integrals of motion there exist a number of extra, particular integrals. It seems natural
to assume that in the quantum case $\bf P=0$ it must occur special properties. It suggests to consider separately the quantum neutral system $\bf {P} = 0$ (the system at rest), where center-of-mass dynamics is absent. It is a goal of the Section.

\hskip 1cm
Substituting $\Psi^{\prime}_{{}_{\mathbf P}}$ into (\ref{HK}) we obtain the equation describing the relative motion
\begin{equation}
 \bigg[ \frac{ {( \mathbf {P}-2\,e\,\mathbf A_{\boldsymbol \rho} )}^2}{2\,M} +
 \frac{{({\mathbf {\hat p}}-e\,(\mu_2-\mu_1)\,{\mathbf A_{\boldsymbol \rho}})}^2}{2\,m_{r}}  -\frac{e^2}{\rho} \bigg]\,\psi_{{}_{\mathbf P}}(\boldsymbol \rho)\ = E\ \psi_{{}_{\mathbf P}}(\boldsymbol \rho)\ .
\label{HKa}
\end{equation}
where CMS momentum $\bf P$ appears as a parameter. At $\bf P=0$, the relative angular momentum $\hat l_z= - i \,\pa_{\varphi}$ is conserved, the relative angle is separated out and the problem (\ref{HK}) is reduced to a study of dynamics in (relative) radial direction, thus, becomes effectively one-dimensional!

\hskip 1cm
At $\bf P=0$ the gauge rotated eigenfunctions (\ref{psik}) (see also (\ref{psiprime})) do not depend on CMS coordinates
\begin{equation}
 \Psi^{\prime}_{{0}}(\bf R\,, \boldsymbol \rho)\ =\ \psi_{{}_{0}}(\boldsymbol \rho)\ .
\label{psirel}
\end{equation}
Furthermore, they admit a factorization in relative polar coordinates ${\boldsymbol \rho}=(\rho,\,\varphi)$ taking the form
\begin{equation}
\psi_{{}_{0}}({\boldsymbol  \rho}) \ =\ \text{e}^{i\,s\,\varphi}\ \text{e}^{-\frac{e\,B}{4\,c}\,\rho^2}\ \rho^{|s|}\ p(\rho)\,,
\label{psi}
\end{equation}
(see \cite{ET-q}), where $s=0,\,\pm 1,\,\pm 2,...$ is the magnetic quantum number. The function $p(\rho)$ obeys
\begin{equation}
\begin{aligned}
&   - \frac{d^2}{d\rho^2}\,p  + \bigg(\frac{e\,B}{c}\,\rho - \frac{1+2\,|s|}{\rho}\bigg)\,\frac{d}{d\rho}\,p + 2\bigg(m_r\,\hat E - \frac{e^2\,m_r}{\rho}\bigg)\,p  \ = \ 0 \ ,
\label{eq}
\end{aligned}
\end{equation}
where, for the sake of convenience, the reference point for the energy is changed,
\[
\hat E \ \equiv \ \frac{e\,B}{2 m_r\,c}\bigg(1+ |s|-|\mu_2-\mu_1|\,s\bigg) - E \ .
\]
The boundary conditions for (\ref{eq}) are chosen in such a way that
\[
   p(\rho) \rar \mbox{const}\quad \mbox{at}\quad \rho \rar 0\quad ,\quad p(\rho) \text{e}^{-\frac{e\,B}{4\,c}\,\rho^2}\ \rho^{|s|} \rar 0\quad \mbox{at}\quad \rho \rar \infty\ .
\]
It is evident that $\hat E =\hat E(e,\,m_1,\,m_2,\,B,\,s)$. For
specific values of a magnetic field $B_n$, the equation (\ref{eq}) possesses polynomial solutions $p=P_n(\rho),\,n=1,2,3..$ \cite{ET-q}. In particular, the largest magnetic field $B_{max}$
for which the problem (\ref{eq}) admits an analytical solution is
\[
 B_{max}\ =\ B_1\ =\ B_0 \equiv 4\,m_r^2\,e^3\,c    \ ,
\]
where $B_0$ is the characteristic magnetic field which defines the magnetic field unit. The eigenfunction reads
\begin{equation}
P_1 \ = \ 1 - \sqrt{\frac{e\, B_1}{c}}\,\rho \ ,
\label{psi1}
\end{equation}
(cf. (\ref{psi1a})),
which corresponds to the first excited state (at magnetic quantum number $s=0$) with the energy $E = E_1 \equiv \frac{e\,B_1}{c\,m_r}$. In general, all known analytically eigenfunctions correspond to excited states. The existence of analytic eigenfunctions is related to the appearance of extra (particular) integrals of motion which happened for a certain values of a magnetic field \cite{ET-q}.

Let us denote an analytic eigenfunction as $\psi^{qes}(\boldsymbol \rho)$. It implies that
\[
{\cal {\hat H}}^{\prime}\,\psi^{qes}\ =\ \frac{e\,B}{2\,m_r\,c}\,\bigg(n+1+|s| - |\mu_2 - \mu_1|\,s \bigg)\,\psi^{qes}\ ,
\]
for a certain quantum numbers $n,s$.
Construct the operator
\begin{equation}
{I}_{n,|s|} \ = \  \tau\, \prod_{j=0}^n(\rho\,{\pa}_\rho-j)\,\tau^{-1}   \ \equiv \  \prod_{j=0}^n(\rho\,{\cal D}_\rho-j)\ ,
\label{Inq}
\end{equation}
where the gauge factor $\tau =  \text{e}^{-\frac{e\,B}{4\,c}\,\rho^2}\ \rho^{|s|}$ and the covariant derivative
\[
{\cal D}_{\rho} = \partial_{\rho} + \frac{e\,B\,\rho}{2\,c} - \frac{|s|}{\rho}\ .
\]
This operator annihilates $\psi^{qes}$,
\[
  {I}_{n,|s|}\,\psi^{qes}\ =\ 0    \ .
\]
Hence, $\psi^{qes}$ is zero mode of ${I}_{n,|s|}$. It immediately implies that
\begin{equation}
[{\cal {\hat H}}^{\prime}\,,\, {I}_{n,|s|} ]\, \psi^{qes}=0\ .
\label{inteq}
\end{equation}
It is evident that in general the operator ${I}_{n,|s|}$ has $(n+1)$ zero modes and for each of them the equation (\ref{inteq}) holds.
Thus, the commutator $[{\cal {\hat H}}\,,\, {I}_{n,|s|}]$ vanishes on the space of zero modes of ${I}_{n,|s|}$. Therefore, ${I}_{n,|s|}$ is
a particular integral (for discussion see \cite{Turbiner:2013}). Moreover, acting over the space of zero modes of ${I}_{n,|s|}$ the neutral system under consideration becomes completely integrable.

\textbf{Scaling relations}

Let us consider two neutral ($q=0$) systems, ($e,\,m_1,\,m_2$) and
($\tilde e,\,\tilde m_1,\,\tilde m_2$). Making scale transformation in (\ref{eq}),
a certain relation between eigenstates emerges. It is easy to check that if
\[
a\ =\ \frac{\tilde e^2\,\tilde m_r}{e^2\,m_r}   \ ,
\]
and
\[
\tilde B \ =\ \frac{\tilde e^3\,\tilde m_r^2}{e^3\,m_r^2}\,B   \ ,
\]
then the following scaling relations occur
\[
\psi(\tilde e,\,\tilde m_1,\,\tilde m_2,\,\tilde B,\,s\,;\,a\rho) \ =\
\psi(e,\,m_1,\,m_2,\,B,\,s;\,\rho) \ ,
\]
\begin{equation}
\label{sc-relq}
\frac{1}{\tilde e^4 \, \tilde m_r}\,
\hat E(\tilde e,\,\tilde m_1,\,\tilde m_2,\,\tilde B,\,s)
 \ = \
\frac{1}{e^4 \,m_r}\,  \hat E(e,\,m_1,\,m_2,\,B,\,s)
  \ .
\end{equation}

\textbf{Asymptotics}

In the case of a neutral system, ($e,\,m_1,\,-e,\,m_2$), making analysis of Eq. (\ref{eq}),
one can obtain that
\begin{equation}
      p\ = \ 1 +c_1\, \rho + c_2\,\rho^2 + \ldots \ ,\quad \rho \rar 0\ ,
\label{rhoq+a}
\end{equation}
where
\[
      c_1 \ = \   -\frac{2\,e^2\,m_r}{1+2\,|s|}                                \ ,
\]
\[
      c_2 \ = \    \frac{1}{2\,(4\,s^2 + 6\,|s| + 2  )}
      \bigg[ 4\,e^4\,m_r^2 + (1+2\,|s|)\bigg(  \frac{e\,B\,(1+|s|- \frac{|m_2-m_1|}{M}\,s  )}{c}  -2\,E\,m_r \bigg) \bigg]                               \ ,
\]
(cf. (\ref{rho+a})), which is, in fact, the perturbation theory expansion near the minimum of the funnel-type potential (see a discussion above). From another end, the expansion at large $\rho \rightarrow \infty $ (WKB asymptotics)
has the form
\begin{equation}
 p\ = \ \rho^\beta(1 + \frac{C_1}{\rho} + \frac{C_2}{\rho^2}\ +\ \ldots)\ ,
 \quad \beta\ =\ \frac{2\,c\,E\,m_r}{B\,e} +\frac{|m_2-m_1|}{M}\,s -|s| -1  \ ,
\label{rhoq+b}
\end{equation}
where
\[
      C_1 \ =    -\frac{2\,c\,e\,m_r}{B}        \ ,
\]
\begin{equation}
\begin{aligned}
      C_2 \ & = \   - \frac{c}{2\,B^3\,e^3\,M^2}\bigg[ 4\,c^2\,E^2\,m_1^2\,m_2^2 - 4\,B\,c\,e\,m_1\,m_2( e^4\,m_1\,m_2 + E\{ M-|m_2-m_1|s     \} )
     \\ & +  B^2\,e^2(M^2 -2\,M\,|m_2-m_1|  s  - 4\,m_1\,m_2\,s^2           )    \bigg]              \ .
\end{aligned}
\end{equation}
(cf. (\ref{rho+b})).

\hskip 1cm
It is worth noting that making analysis of (\ref{HK}) one can find a behavior of
the ground state energy $E_0$ at weak and strong magnetic fields. For the weak-magnetic-field limit one can be derived using perturbation theory in powers of $B^2$ in (\ref{HK}). Final result has a form
\begin{equation}
E_0\ =\ -2\,m_r\,e^4 +  \frac{3}{64\,c^2\,e^2\,m_r^3}\,B^2 + \ldots \ .
\label{weak}
\end{equation}
It seems evident that this series is asymptotic. For the high-magnetic-field limit, making a suitable rescaling of $\rho$ coordinate in (\ref{HK}) and developing a perturbation theory with respect to the Coulomb interaction term in the potential, we arrive at
\begin{equation}
E_0\ =\ \frac{e\,B}{2\,c\,m_r}\, -\,\sqrt{\frac{e^5\,\pi}{2\,c}}\ B^{1/2}\, +\, \ldots \ .
\label{strong}
\end{equation}

\subsection{Approximations}

Following the same strategy as for the case $e_c=0$, see Section II, we make an interpolation for the eigenfunction $p$ in (\ref{eq}) between perturbation theory at small distance (\ref{rhoq+a}) and the WKB expansion at large distances (\ref{rhoq+b}) also keeping in mind a form of the exact solutions of (\ref{eq}) (see \cite{ET-q}) which must emerge for specific values of a magnetic field. It is worth mentioning that all exact solutions of (\ref{eq}) correspond to excited states.
It is not that surprising but the simplest interpolation for lowest states have the forms given by (\ref{grounda})-(\ref{2THa}) like for $e_c=0$ case.

\subsection{Results}

\hskip 1cm
We focus on the important particular case of neutral system: the Hydrogen atom.
Other neutral systems, $q=0$, can be studied using the scaling relation (\ref{sc-relq}). The obtained results for several low-lying states $(n,s)$ with relative quantum numbers $n=0,1,2$ and relative magnetic quantum numbers $s=0,1,2$ are presented in Tables \ref{Table6} - \ref{Table10}.

\begin{figure}[htp]
\begin{center}
\includegraphics[width=3.5in,angle=-90]{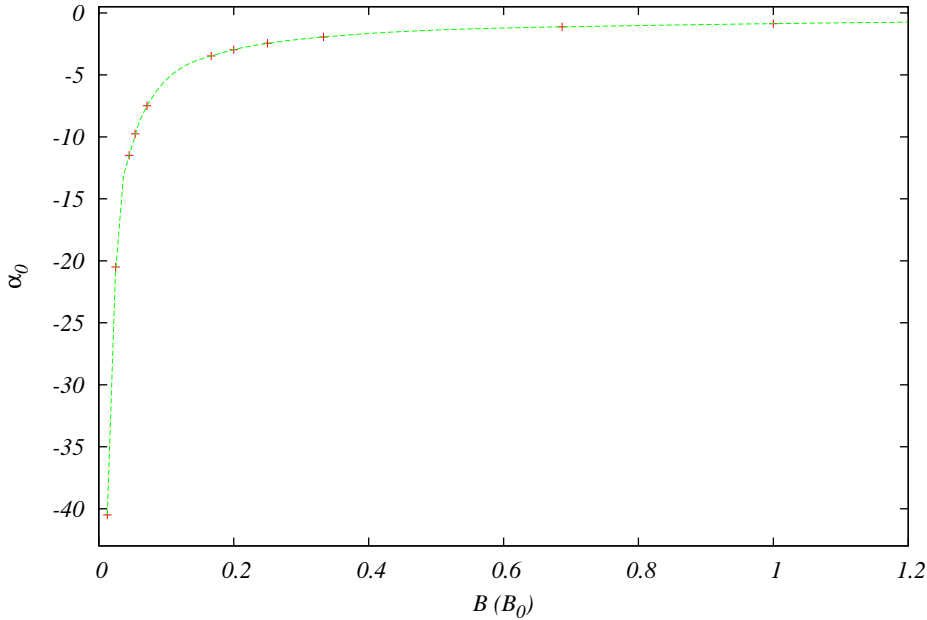}
\caption{Ground state for the $2D$ Hydrogen atom: optimal value of the parameter $\al_0$ in (\ref{grounda})
{\it vs} magnetic field $B$\ .}
\label{q-alpha0}
\end{center}
\end{figure}

An immediate observation is that the very simple, few-parametric, variational trial functions (\ref{grounda}), (\ref{1THa}), (\ref{2THa}) lead to highly accurate variational energies, see Tables \ref{Table6}, \ref{Table7}, \ref{Table9} as well as position of nodes, see Tables \ref{Table8}, \ref{Table10}. Optimal parameters depend smoothly on a magnetic field slowly changing with a magnetic field variation. On Fig.\ref{q-alpha0} the behavior of the parameter $\al_0$ in (\ref{grounda}) is shown. At large $B$ the optimal $\al_0$ monotonically tends to zero.

For weak magnetic fields the energies decrease linearly with a magnetic field decrease tending to zero (cf. (\ref{weak})). Similar linear behavior of the energy appears at large magnetic fields: it grows linearly with a magnetic field increase (cf. (\ref{strong})). For the ground state energy, in the Born-Oppenheimer approximation, $m_1 \rar \infty $, one can compare our calculations with those carried out in the so called Asymptotic Iteration Method \cite{Soy}, see Table \ref{qComparison}: our variational energies are systematically lower.

In a similar way as was done for $e_c=0$ case one can pose a question about the accuracy
of obtained variational results, in particular, how a variational trial function is close to the exact one. The answer to the question is given in a framework of a convergent perturbation theory \cite{Turbiner:1979-84} (see Appendix A) where the variational trial function is taken as zero approximation. It allows us to estimate a deviation of the variational trial function from the
exact one.

\begin{figure}[htp]
\centering
\subfigure[]{\includegraphics[width=2.5in,angle=0]{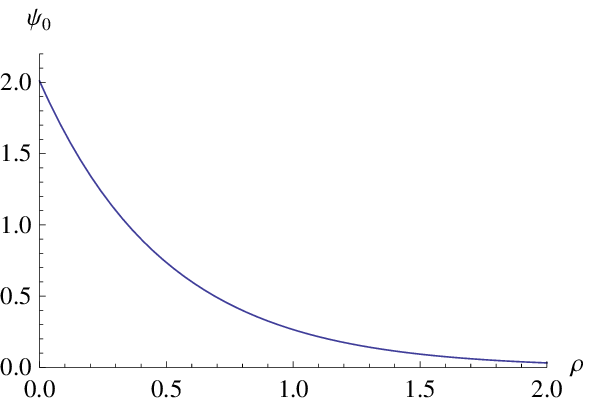}} \qquad \subfigure[]{\includegraphics[width=2.5in,angle=0]{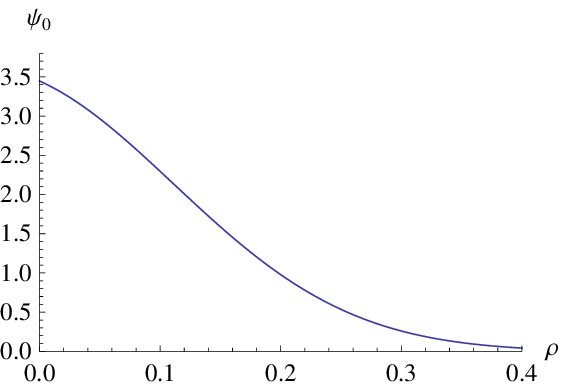} }
\vspace{0.5cm} \\
\subfigure[]{\includegraphics[width=2.5in,angle=0]{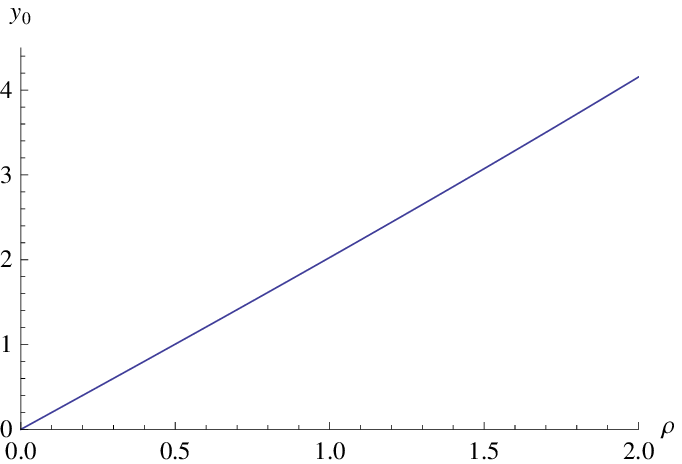}} \qquad \subfigure[]{\includegraphics[width=2.5in,angle=0]{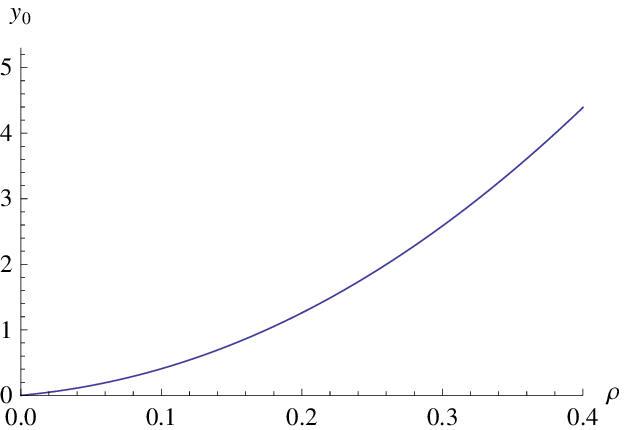} }
\vspace{0.5cm} \\
\subfigure[]{\includegraphics[width=2.6in,angle=0]{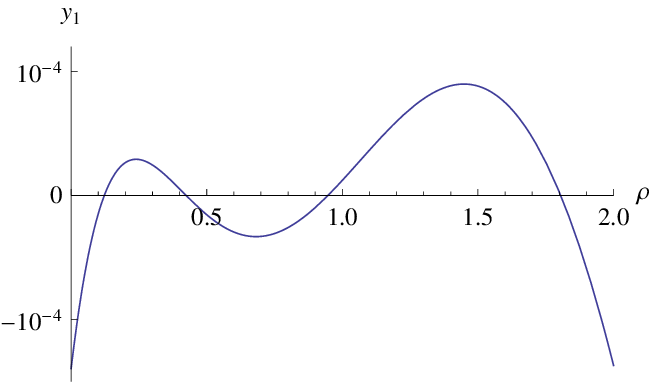}} \qquad \subfigure[]{\includegraphics[width=2.6in,angle=0]{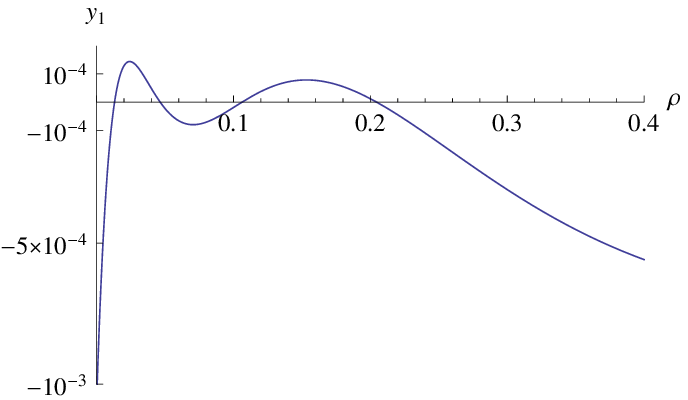} }
\caption{Ground state of Hydrogen atom at $s=0$, at $B=\frac{1}{5}B_0$ (left)
and $B=25\,B_0$ (right):
(a)-(b) eigenfunction (\ref{psia}), with prefactor (\ref{grounda}),
taken as zero approximation $\psi \equiv e^{-\phi_0}$,
(c)-(d) $y_0(\rho) = (\phi_0)'$ and (e)-(f) the first correction
$y_1(\rho)$ (see (\ref{yn}))\ .\
$B_0 = 9.391766\times 10^9 \,G$.}
\label{y1q}
\end{figure}

Analysis of the first correction to the eigenfunction allows us to draw a conclusion that the trial functions (\ref{grounda}), (\ref{1THa}), (\ref{2THa}) at optimal values of
parameters are very accurate uniform approximations of the exact eigenfunction. Locally, the approximation provides at least 3-5 significant digits (s.d.)
exactly for any value of the external magnetic field strength(!) (see Fig. \ref{y1q}). Furthermore, for a domain which gives a dominant contribution to energy integral, $\langle \psi_{trial} {\cal H} \psi_{trial} \rangle$ the number of s.d. increases to 9-10. This is the reason why the variational energy gets known with 8-10 significant digits.

\begin{figure}[htp]
\begin{center}
\includegraphics[width=3.5in,angle=-90]{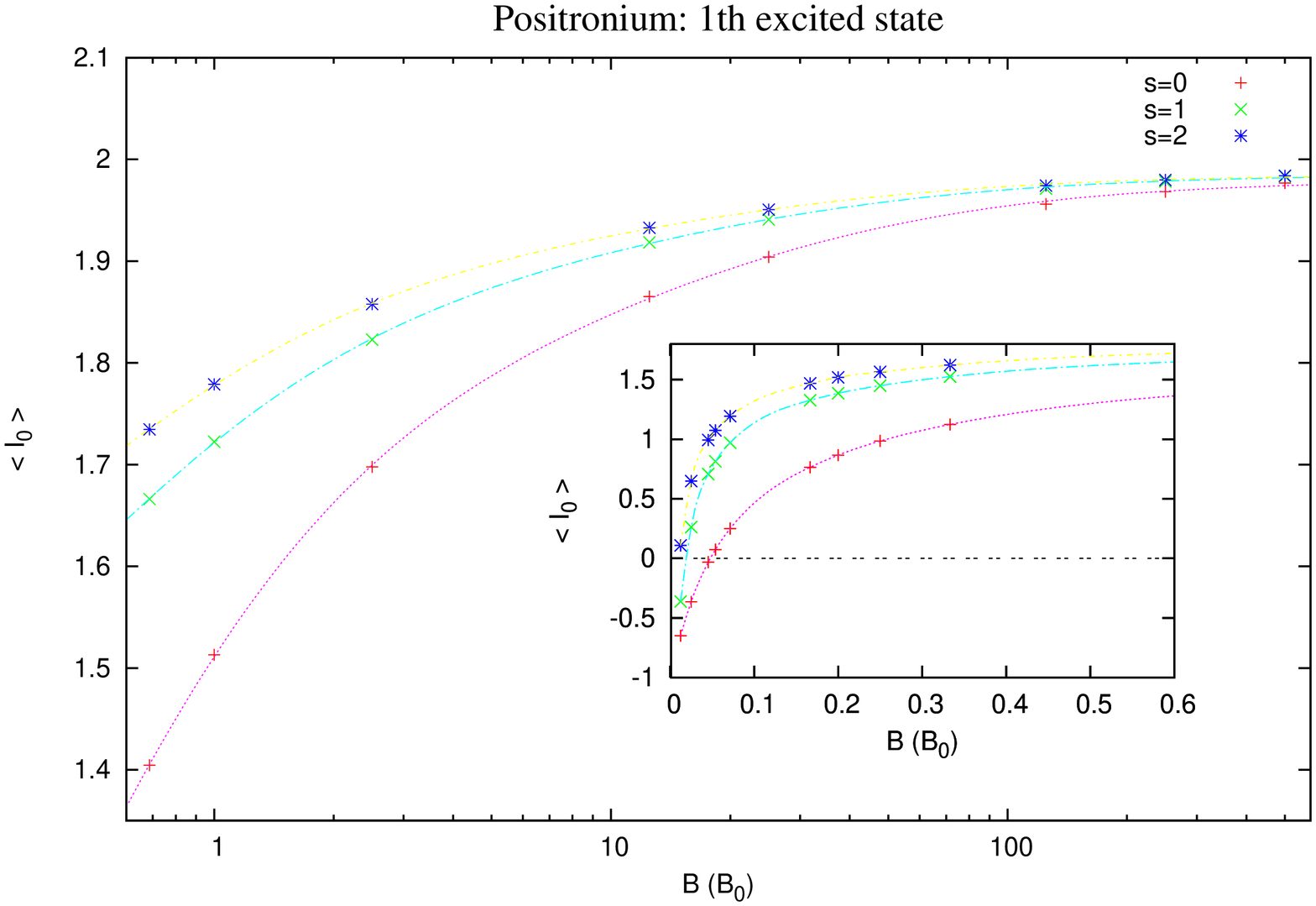}
\caption{Expectation value $\langle {I}_{0,|s|} \rangle$, see (\ref{Inq}) at $s=0,1,2$
{\it vs} magnetic field $B$, $B_0 = 2.3505  \times 10^9 G\ .$
\label{Fig.3}}
\end{center}
\end{figure}

\begin{figure}[htp]
\begin{center}
\includegraphics[width=3.5in,angle=-90]{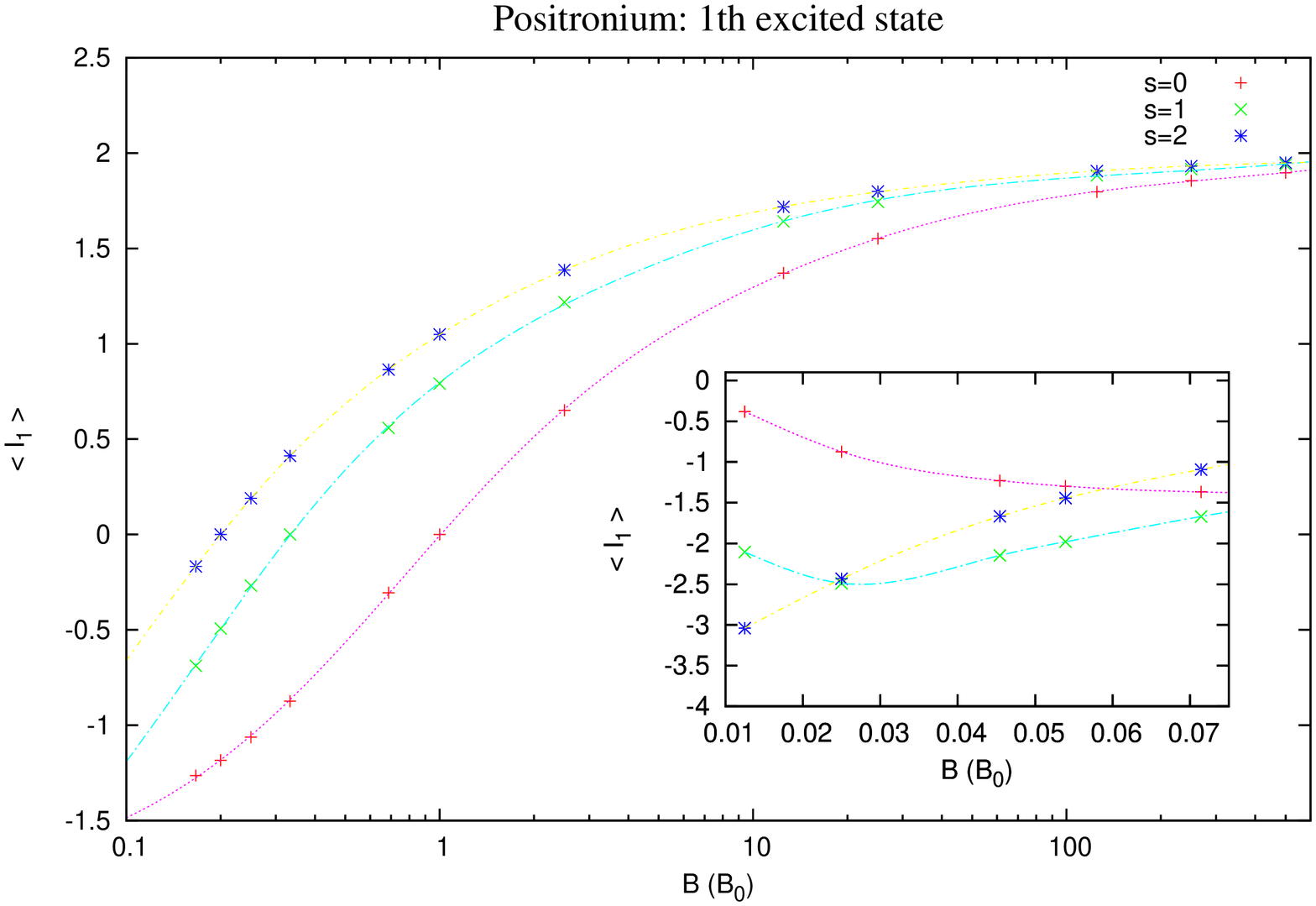}
\caption{Expectation value $\langle {I}_{1,|s|} \rangle$, see (\ref{Inq}) at $s=0,1,2$
{\it vs} magnetic field $B$. For $B=\frac{1}{1+2\,|s|}\ B_0 $, the analytic solutions
in (\ref{eq}) occur at $n=1$ and $\langle {I}_{1,|s|} \rangle$ vanishes,
$B_0 = 2.3505 \times 10^9 G\ .$
\label{Fig.4}}
\end{center}
\end{figure}

In the case of Positronium, on Figs. \ref{Fig.3} - \ref{Fig.4} the expectation values for the particular integrals ${I}_{0,|s|}$ and ${I}_{1,|s|}$ at $s=0,1,2$, see (\ref{Inq}), are shown. Notice that at $B_s=\frac{1}{1+2\,|s|}\ B_0$ the expectation value $\langle {I}_{1,|s|} \rangle$ vanishes. It corresponds to the appearance of the analytic solution in (\ref{eq}) at $n=1$,
${I}_{1,|s|}\,\psi^{qes}=0$.

\section{Conclusions}

Summarizing, we state that a simple uniform approximation of the lowest eigenfunctions for two particular, physically important quantum systems, $e_c=0$ and $q=0$ at rest is presented.  It manifests an approximate solution of the problem of spectra of this systems.
The key element of the procedure is to construct an interpolation between the WKB expansion at large distances and perturbation series at small distances for the phase of the wavefunction, or, in other words, to find an approximate solution for the corresponding eikonal equation. Separation of variables helps us to solve this problem. It is interesting that for both systems
there exists special discrete set of magnetic fields for which some observables take values which can be found exactly, the eigenfunctions and energies of some states are known explicitly in closed analytic form.

For moving neutral system along the plane in presence of the magnetic field CMS motion is factored out although CMS variables are not separated. Factorization in the space of the relative coordinates is absent. As a result the WKB asymptotics cannot be constructed in a uniform way, it depends on a way of approaching to infinity. However, a reasonable approximation of the first growing terms of the WKB expansion seems sufficient to construct the interpolation between large and small distances giving high accuracy results.
It will be done elsewhere \cite{ET-P:2013}.

\begin{acknowledgments}
  The authors are grateful to J. C. L\'opez Vieyra and H. Olivares Pilon for their interest in the present work, helpful discussions and important assistance with computer calculations.
  This work was supported in part by the University Program FENOMEC and by the PAPIIT
  grant {\bf IN109512} and CONACyT grant {\bf 166189}~(Mexico).
\end{acknowledgments}

\pagebreak

\vspace{0.3cm}

{\bf Appendix A.\ Perturbation Theory}

\vspace{0.2cm}

After performing variational study a natural question to ask concerns the accuracy of obtained results. In particular, how close a trial function is to exact one, how accurate the variational energy. In order to answer the question we develop a convergent perturbation theory in the Schr\"{o}edinger equation with respect to the deviation of the original potential to the trial potential (see below).

Let us take the radial Schr\"{o}edinger equation
\begin{equation}
(\De_{\rho} + V - E ) \, \psi \ =\ 0\ ,
\label{SchroEq-ec}
\end{equation}
where $\Delta_{\rho}=\frac{d^2}{d\rho^2}+\frac{1}{\rho}\frac{d}{d\rho}$, $V=V(\rho)$ and construct the perturbation theory of the so-called \textit{non-linearization procedure} \cite{Turbiner:1979-84}. Assume we choose some trial,
nodeless function $\psi_0(\rho)$. By definition it can be considered as the ground state function in the potential, $\frac{\De_{\rho}\, \psi_0}{\psi_0} \equiv V_0$ with the zero ground state energy, $E_0=0$. Immediately, we can write the original potential as $V = V_0 + \la\,V_1$ at $\la=1$ and $V_1=V-V_0$. Now one can develop perturbation theory in powers of $\la$,
\begin{equation}
\label{PT0}
     E=\sum \la^j E_j\quad ,\quad \psi = \psi_0\,\text{e}^{-\sum \la^j \phi_j}\ .
\end{equation}
As a first step let us transform (\ref{SchroEq-ec}) into the Riccati equation
form (assuming for simplicity $m_r=\frac{1}{2}$) by introducing $\psi = {\rm e}^{-\phi}$,
\begin{equation}
 y^\prime -y^2+\frac{y}{\rho} \ = \ E - V_0 - \la V_1   \ , \qquad y=\phi^\prime \ .
\label{Riccati}
\end{equation}
At $\la=0$ the solution of (\ref{Riccati}) is given by $y_0=(\log \psi_0)^{\prime}$ and $E_0=0$.
It is easy to find the equation for $j$th correction $y_j = (\phi_j)^{\prime}$,
\begin{equation}
y_j^\prime - (2\,y_0 - \frac{1}{\rho} )y_j \ = \ E_j - V_j\ ,
\label{corr}
\end{equation}
where $V_j=\sum_{i=1}^{j-1}y_i\,y_{j-i}$ for $j>1$. Its solution has a form
\begin{equation}
y_j\ =\ \frac{1}{\rho\,\psi^2_0(\rho)}\int^{\rho}_0 (E_j-V_j(x))\ \psi^2_0(x)\,x\,dx \ ,
\label{yn}
\end{equation}
and
\begin{equation}
E_j\ =\ \frac{\int^{\infty}_0\, V_j\ \psi^2_0(\rho)\,\rho\,d\rho   }{\int^{\infty}_0\psi^2_0(\rho)\, \rho\,d\rho  }  \ ,
\label{En0}
\end{equation}
as a consequence of the boundary condition,
\[
 y_j\, \rho\,\psi^2_0(\rho) \rar 0\ \mbox{at}\ \rho \rar \infty \ .
\]

In the case of excited (nodefull) states the perturbation theory is modified. At first, for
the $n$th excited state in $\rho$ with magnetic quantum number $s$ the eigenfunction can be taken in the representation
\begin{equation}
\label{psi0n}
      \psi^{(n)}(\rho)\ =\ \rho^{|s|} \ \prod_{i=1}^{n} (\rho - f^{(i)})\ e^{-\phi}\ ,
\end{equation}
where $f^{(i)},\ i=1,\ldots n$ are nodes. Then, the perturbation theory is developed separately for energy, phase
\[
     E\ =\ \sum \la^j E_j\quad ,\quad \psi\ =\ \text{e}^{-\sum \la^j \phi_j}\ ,
\]
and nodes,
\begin{equation}
\label{PTn}
     f^{(i)}\ =\ \sum \la^j f^{(i)}_j \ ,
\end{equation}
(cf. (\ref{PT0})). The explicit formulas for corrections can be derived (cf. \cite{Turbiner:1979-84}). In particular, for the first excited state the first correction to a node
$f^{(1)}$,
\[
f^{(1)}_1\ =\ \frac{1}{\rho_0 e^{-2 \phi_0(\rho_0)}} \int_0^{\rho_0} ( V_1 - E_1 ) \psi^2_0(\rho)\, \rho\,d\rho \quad ,\quad \rho_0=f^{(1)}_0\ ,
\]

\hskip 1cm
A sufficient condition for such a perturbation theory to be convergent is to require that a perturbation `potential' has to be bounded:
\[
 |V_1(\rho)| < C \ ,
\]
where $C$ is constant. Obviously, the rate of convergence gets faster for smaller values of $C$~. It is evident that the perturbation
$V_1(\rho)$ is bounded if $\phi_0(\rho)$ is a smooth function which reproduces exactly all growing terms at $\rho$ tending to infinity including
the logarithmic term $\sim \beta \log \rho$ stemming from the expansion (\ref{rho+b}). Eventually, we choose $\phi_0$ as in (\ref{psia})
multiplied by a factor (\ref{grounda}), (\ref{1THa}), (\ref{2THa}), respectively, which generates the logarithmic term in the expansion of the phase at $\rho \rar \infty$. Hence,  the emerging perturbation theory has to be convergent.

It was shown in \cite{Turbiner:1979-84} the variational energy calculated with a trial function $\psi_{trial}$ is equal to sum of the first two terms of the perturbation theory where $\psi_{trial}$ is taken as zero approximation, $E_{variational}=E_0 + E_1$. If this perturbation theory with $\psi_{0}=\psi_{trial}$ is convergent, then the first correction $\psi_{1}$ to the trial function characterizes the deviation from the exact eigenfunction, while $E_2$ gives an estimate of the accuracy of the variational energy.

\clearpage

\begin{center}
\textbf{Two electron case $(e_c=0)$. Energy of the relative motion $E_\rho$}
\end{center}
\setlength{\tabcolsep}{18.0pt}
\setlength{\extrarowheight}{1.0pt}
\begin{table}[th]
\begin{center}
\begin{tabular}{|c||c|c|c|}
\hline
\\[-23pt]
$B$ ($B_{{}_0}$)  & \multicolumn{3}{c|}{ $E_\rho$ } \\
 \hline                 & \multicolumn{3}{l|}{\hspace{1.0cm} $s=0$ \hspace{3.0cm} $s=1$ \hspace{3.6cm} $s=2$ }\\
\hline
\hline
$\frac{1}{80}$                           &   $0.0748326441(3)$       &   $0.0642857285(2)$       &   $0.05697827291(4)$
\\[3pt]
\hline
$\frac{1}{40}$                            &   $0.12313258158(5)$      &   $0.1031020(7)$          &   $0.090532124234(3)$
\\[3pt]
\hline
$\frac{1}{22}$                            &   $0.19027307879(3)$      &   $0.155914269(3)$        &   $0.13636363636363^*$
\\[3pt]
\hline
$\frac{1}{10+\sqrt{73}}$                  &   $0.2157031488418^*$     &   $0.17572661(6)$         &   $0.1536438753898(7)$
\\[3pt]
\hline
$\frac{1}{14}$                            &   $0.26541068608(4)$      &   $0.2142857142(7)$       &   $0.187419053899(5)$
\\[3pt]
\hline
$ \frac{1}{6}$                            &   $0.5^* $                &   $0.395527795(8)$        &   $0.3484374354981(3)$
\\[3pt]
\hline
$\frac{1}{5}$                             &   $0.573970670(3)$        &   $0.45284939(5)$         &   $0.4^*$
\\[3pt]
\hline
$\frac{1}{4}$                             &   $0.6801405(7)$          &   $0.53543795(4)$         &   $0.4747145722810(7)$
\\[3pt]
\hline
$\frac{1}{3}$                             &   $0.84773000(8)$         &   $\frac{2}{3}^*$         &   $0.594292534436(5)$
\\[3pt]
\hline
$\frac{1}{10-\sqrt{73}}$                  &   $1.48626025(8)$         &   $1.17648943(2)$         &   $1.065706008677(1)$
\\[3pt]
\hline
$1$                                       &   $ 2.0^*  $              &   $1.596498881(6)$        &   $1.45929137837(3)$
\\[3pt]
\hline
$\frac{5}{2}$                             &   $4.19942226(0)$         &   $3.459757282(1)$        &   $3.23226037856(0)$
\\[3pt]
\hline
$\frac{25}{2}$                            &   $16.616100476(1)$       &   $14.683577109(5)$       &   $14.150653782104(0)$
\\[3pt]
\hline
$25$                                      &   $30.942950650(3)$       &   $28.1010717338(2)$      &   $27.338904226984(3)$
\\[3pt]
\hline
$125$                                     &  $138.676904357(0)$       &   $131.97372337706(2)$    &   $130.2435683148197(5)$
\\[3pt]
\hline
$250$                                     &  $269.478147780(8)$       &   $259.8757303773(5)$     &   $257.4201155478(7)$
\\[3pt]
\hline
$500$                                     &  $527.6843872(2)$         &   $513.97983895(2)$       &   $510.49822676599(4)$
\\[3pt]
\hline
\end{tabular}
\caption{\small Ground state energy $E_\rho$ in Hartrees (see (\ref{Psi1})), a modification due to correction $E_2$ (\ref{En0})
indicated by a number in brackets; magnetic field in effective atomic units, $B_0 = 4.701\times 10^9 \,G$. Energies corresponding to exact solutions marked by *\,.}
\label{Table1}
\end{center}
\end{table}

\clearpage

\begin{center}
\textbf{Two electron case $(e_c=0)$. Energy of the relative motion $E_\rho$}
\end{center}

\setlength{\tabcolsep}{10.0pt}
\setlength{\extrarowheight}{2.0pt}
\begin{table}[th]
\scriptsize
\begin{center}
\begin{tabular}{|c||c|c|c|}
\hline
\\[-16pt]
$B$ ($B_{{}_0}$)  & \multicolumn{3}{c|}{ $E_\rho$ } \\
\hline               & \multicolumn{3}{l|}{ \hspace{0.5cm} $s=0$ \hspace{2.1cm} $s=1$ \hspace{2.3cm} $s=2$  }\\
\hline
\hline
$\frac{1}{80}$                            &   $0.09699(2)$      &  $0.08671439(5)$        &  $0.079897218(2)$
\\[3pt]
\hline
$\frac{1}{40}$                            &   $0.16765(4)$      &  $0.14841894(0)$        &  $0.1370070010(2)$
\\[3pt]
\hline
$\frac{1}{22}$                            &   $0.27156(1)$      &  $0.23916872(8)$        &  $0.2218773676(5)$
\\[3pt]
\hline
$\frac{1}{10+\sqrt{73}}$                  &   $0.31225(8)$      &  $0.2748122(8)$         &  $0.2554304488(1)$
\\[3pt]
\hline
$\frac{1}{14}$                            &   $0.39356(7)$      &  $0.346241372(2)$       &  $0.32295527422(2)$
\\[3pt]
\hline
$ \frac{1}{6}$                            &   $0.80103(3)$      &  $0.708595501(3)$       &  $0.66922392903535(3)$
\\[3pt]
\hline
$\frac{1}{5}$                             &   $0.935800(5)$     &  $0.8298538863(2)$      &  $0.785980199872(5)$
\\[3pt]
\hline
$\frac{1}{4}$                             &   $1.133400(7)$     &  $1.00868104442(1)$     &  $0.958687644351(5)$
\\[3pt]
\hline
$\frac{1}{3}$                             &   $1.4539363(1)$    &  $1.3009640149(4)$      &  $1.24197319573(7)$
\\[3pt]
\hline
$\frac{1}{10-\sqrt{73}}$                  &   $2.7472598141^*$  &  $2.4990089725(0)$      &  $2.41064828188(4)$
\\[3pt]
\hline
$1$                                       &   $3.84698944(5)$   &  $3.532376896(3)$       &  $3.42395547396(2)$
\\[3pt]
\hline
$\frac{5}{2}$                             &   $8.8931012(6)$    &  $8.350874458(0)$       &  $8.17427228881(6)$
\\[3pt]
\hline
$\frac{25}{2}$                            &   $40.732242(5)$    &  $39.422380852(1)$      &  $39.0162273893(0)$
\\[3pt]
\hline
$25$                                      &   $79.6096972(1)$   &  $77.725353318(1)$      &  $77.1471465895(1)$
\\[3pt]
\hline
$125$                                     &  $385.4205625(3)$   & $381.114147747(2)$      &  $379.80978632347(0)$
\\[3pt]
\hline
$250$                                     &  $764.774066(8)$    & $758.653457765(8)$      &  $756.80496202143(4)$
\\[3pt]
\hline
$500$                                     & $1520.9305898(5)$   & $1512.244591054(3)$     &  $1509.62656921827(3)$
\\[3pt]
\hline
\end{tabular}
\caption{\small First excited state energy $E_\rho$ in Hartrees (see (\ref{Psi1})), a
modification due to correction $E_2$ (\ref{En0}) indicated by a number in brackets; magnetic field in effective atomic units,
$B_0 = 4.701\times 10^9 \,G$. Energies corresponding to exact solutions marked by *\,.}
\label{Table2}
\end{center}
\end{table}

\clearpage

\begin{center}
\textbf{Two electron case $(e_c=0)$. Node for first excited state}
\end{center}

\setlength{\tabcolsep}{12.0pt}
\setlength{\extrarowheight}{2.0pt}
\begin{table}[th]
\begin{center}
\begin{tabular}{|c||c|c|c|}
\hline
\\[-24pt]
 $B$ ($B_{{}_0}$)  & \multicolumn{3}{c|}{$s=0$ \hspace{1.5cm} $s=1$ \hspace{1.9cm} $s=2$  } \\
   \hline                & \multicolumn{3}{l|}{\hspace{0.3cm} $f_0 +f_1$  \hspace{1.5cm}  $f_0 +f_1$ \hspace{1.8cm}  $f_0 +f_1$ }\\
\hline
\hline
$\frac{1}{80}$                            &   $25.90(9)$     &   $26.58(2)$         &   $28.176(4)$
\\[3pt]
\hline
$\frac{1}{40}$                            &   $16.70(7)$     &   $17.369(6)$        &   $18.777(4)$
\\[3pt]
\hline
$\frac{1}{22}$                            &   $11.4(8)$      &   $12.127(6)$        &  $13.3516(6)$
\\[3pt]
\hline
$\frac{1}{10+\sqrt{73}}$                  &   $10.3(2)$      &   $10.959(8)$        &   $12.129(4)$
\\[3pt]
\hline
$\frac{1}{14}$                            &   $8.66(5)$      &   $9.292(1)$         &   $10.37153(8)$
\\[3pt]
\hline
$ \frac{1}{6}$                            &   $5.140(8)$     &   $5.71201(6)$       &   $6.529552(2)$
\\[3pt]
\hline
$\frac{1}{5}$                             &   $4.59(9)$      &   $5.1550(4)$        &   $5.920251(6)$
\\[3pt]
\hline
 $\frac{1}{4}$                            &   $4.015(5)$     &   $4.551105(7)$      &   $5.254933(2)$
\\[3pt]
\hline
$\frac{1}{3}$                             &   $3.374(3)$     &   $3.88152(0)$       &   $4.5109(0)$
\\[3pt]
\hline
$\frac{1}{10-\sqrt{73}}$                  &  $2.19215^*$     &   $2.61977(7)$       &   $3.08710(4)$
\\[3pt]
\hline
$1  $                                     &  $1.7586(5)$     &   $2.14286(3)$       &   $2.540094(4)$
\\[3pt]
\hline
$\frac{5}{2}  $                           &  $1.0402(5)$     &   $1.32279(4)$       &   $1.585670(3)$
\\[3pt]
\hline
$\frac{25}{2}  $                          &  $0.4306(8)$     &   $0.577399(7)$      &   $0.7001600(2)$
\\[3pt]
\hline
$25  $                                    &  $0.29835(8)$    &   $0.4058732(1)$     &   $0.4935730(1)$
\\[3pt]
\hline
$125$                                     &  $0.129638(2)$   &   $0.1800643(1)$     &   $0.21982542(2)$
\\[3pt]
\hline
$250   $                                  &  $0.091021(2)$   &   $0.1270810(4)$     &   $0.15528769(7)$
\\[3pt]
\hline
$500  $                                   &  $0.064036(5)$   &   $0.08973786(4)$    &   $0.109728751(2)$
\\[3pt]
\hline
\end{tabular}
\caption{\small First excited state: evolution of node
$f$ in a.u., modification due to the first correction $f_1$ (\ref{PTn}) indicated by a number in brackets. $B_{{}_0}= 4.701\times 10^9 \,G$.}
\label{Table3}
\end{center}
\end{table}

\clearpage

\begin{center}
\textbf{Two electron case $(e_c=0)$. Energy of the relative motion $E_\rho$}
\end{center}

\setlength{\tabcolsep}{12.0pt}
\setlength{\extrarowheight}{2.0pt}
\begin{table}[th]
\scriptsize
\begin{center}
\begin{tabular}{|c||c|c|c|}
\hline
\\[-16pt]
$B$ ($B_{{}_0}$)  & \multicolumn{3}{c|}{ $E_\rho$ } \\
 \hline              & \multicolumn{3}{l|}{ \hspace{0.6cm} $s=0$ \hspace{2.2cm} $s=1$ \hspace{2.4cm} $s=2$  }\\
\hline
\hline
$\frac{1}{80}$                            &   $0.11956(7)$         &  $0.109535(4)$           &   $0.10313548(6)$
\\[3pt]
\hline
$\frac{1}{40}$                            &   $0.21312(5)$         &  $0.1945971(7)$          &   $0.184110892(4)$
\\[3pt]
\hline
$\frac{1}{22}$                            &   $0.35476(4)$         &  $0.3240510(8)$          &   $0.308461(1)$
\\[3pt]
\hline
$\frac{1}{10+\sqrt{73}}$                  &   $0.41115(2)$         &  $0.3758344(6)$          &   $0.358451838(0)$
\\[3pt]
\hline
$\frac{1}{14}$                            &   $0.52496(2)$         &  $0.4807512(4)$          &   $0.460041855(2)$
\\[3pt]
\hline
$\frac{1}{6}$                             &   $1.1106(5)$          &  $1.02716875(5)$         &   $0.99292908(0)$
\\[3pt]
\hline
$\frac{1}{5}$                             &   $1.30820(5)$         &  $1.21327129(4)$         &   $1.175274255(2)$
\\[3pt]
\hline
$\frac{1}{4}$                             &   $1.60027(4)$         &  $1.48960942(5)$         &   $1.446517136(8)$
\\[3pt]
\hline
$\frac{1}{3}$                             &   $2.07891(0)$         &  $1.94488148(7)$         &   $1.89431649(7)$
\\[3pt]
\hline
$\frac{1}{10-\sqrt{73}}$                  &   $4.04885(6)$         &  $3.83779886(5)$         &   $3.762927107(2)$
\\[3pt]
\hline
$1$                                       &   $5.753021(2)$        &  $5.48923820(0)$         &   $5.397803046(8)$
\\[3pt]
\hline
$\frac{5}{2} $                            &  $13.721761(2)$        &  $13.279436628(1)$       &  $13.1317958068(2)$
\\[3pt]
\hline
$\frac{25}{2}$                            &  $65.2941228(0)$       &  $64.255403826(2)$       &  $63.9187675993(7)$
\\[3pt]
\hline
$25$                                      &  $128.9708709(1)$      & $127.486713200(1)$       & $127.0084693311(5)$
\\[3pt]
\hline
$125$                                     &  $633.9344238(3)$      & $630.572977759(0)$       & $629.4971396326(2)$
\\[3pt]
\hline
$250$                                     &  $1262.6530447(5)$     & $1257.885573370(5)$      & $1256.3619510814(0)$
\\[3pt]
\hline
$500$                                     &  $2517.9117379(2)$     & $2511.156076611(4)$      & $2508.9991935583(2)$
\\[3pt]
\hline
\end{tabular}
\caption{\small The second excited state energy $E_\rho$ in Hartrees(see (\ref{Psi1})), a
modification due to correction $E_2$ (\ref{En0}) indicated by a number in brackets; magnetic field
in effective atomic units, $B_0 = 4.701\times 10^9 \,G$. Energies corresponding to exact solutions marked by *\,.}
\label{Table4}
\end{center}
\end{table}

\clearpage

\begin{center}
\textbf{Two electron case $(e_c=0)$. Nodes $f^{a(b)}$ (a.u.) for 2nd-excited state}
\end{center}

\setlength{\tabcolsep}{4.0pt}
\setlength{\extrarowheight}{2.3pt}
\begin{table}[th]
\begin{center}
\begin{tabular}{|c||c|c|c|c|c|c|}
\hline
\\[-24pt]
$B$ ($B_{{}_0}$)  & \multicolumn{6}{c|}{      $s=0$   \hspace{3.8cm}     $s=1$        \hspace{3.8cm}     $s=2$   } \\[2pt]\hline
& \multicolumn{6}{l|}{\hspace{0.5cm} $f^a_0 + f^a_{1}$  \hspace{0.9cm}  $f^b_0 + f^b_{1}$ \hspace{1.1cm}    $f^a_0 + f^a_{1}$  \hspace{0.9cm}  $f^b_0 + f^b_{1}$ \hspace{1.0cm}  $f^a_0 + f^a_{1}$  \hspace{0.9cm}  $f^b_0 + f^b_{1}$ \hspace{0.5cm} }\\[2pt]
\hline
\hline
$\frac{1}{80}$                    &   $   23.2(7)         $    &  $ 34.5(3)       $  & $ 21.6(5)       $    & $ 35.(1)       $    & $ 23.(2)       $  &    $ 36.(6)       $
\\[2pt]
\hline
$\frac{1}{10+\sqrt{73}}$          &   $   8.106(4)        $    &  $ 14.66(4)      $  & $ 8.7842(2)     $    & $ 15.27(2)     $    & $ 9.9443(6)    $  &    $ 16.343(7)    $
\\[2pt]
\hline
$\frac{1}{14}$                    &   $   6.772(8)        $    &  $ 12.4740(7)    $  & $ 7.433(5)      $    & $ 13.0649(7)   $    & $ 8.49582(3)   $  &    $ 14.0483(2)   $
\\[2pt]
\hline
$\frac{1}{4}$                     &   $   3.080(2)        $    &  $  6.13614(8)   $  & $ 3.6239(6)     $    & $ 6.62061(8)   $    & $ 4.29523(3)   $  &    $ 7.247648(1)  $
\\[2pt]
\hline
$\frac{1}{10-\sqrt{73}}$          &   $   1.6672(2)       $    &  $  3.51393(1)   $  & $ 2.08400(4)    $    & $ 3.88641(7)   $    & $ 2.52167(5)   $  &    $ 4.29770(9)   $
\\[2pt]
\hline
$1$                               &   $   1.3354(6)       $    &  $  2.86651(3)   $  & $ 1.70448(8)    $    & $ 3.196815(4)  $    & $ 2.07459(5)   $  &    $ 3.54524(8)   $
\\[2pt]
\hline
$\frac{5}{2} $                    &   $   0.789170(8)     $    &  $  1.75798(5)   $  & $ 1.052268(9)   $    & $ 1.994351(6)  $    & $ 1.294858(6)  $  &    $ 2.22353(7)   $
\\[2pt]
\hline
$25  $                            &   $   0.22726581(9)   $    &  $  0.533587(3)  $  & $ 0.323033(7)   $    & $ 0.620201(9)  $    & $ 0.40300(7)   $  &    $ 0.6961070(5) $
\\[2pt]
\hline
$250  $                           &   $   0.069543019(7)  $    &  $  0.16637(7)   $  & $ 0.10117071(2) $    & $ 0.1950616(7) $    & $ 0.126792(1)  $  &    $ 0.219418(3)  $
\\[2pt]
\hline
$500  $                           &   $   0.048949757(6)  $    &  $ 0.1174175(6)  $  & $ 0.07144426(4) $    & $ 0.1378271(5) $    & $ 0.0895932(8) $  &    $ 0.15508(4)   $
\\[2pt]
\hline
\end{tabular}
\caption{Modification of node position due to correction $f_1$ (\ref{PTn}) indicated by a number in brackets.}
\label{Table5}
\end{center}
\end{table}

\clearpage


\clearpage

\begin{center}
\textbf{Hydrogen atom $(q=0)$ at $\mathbf P=0$. Total energy $E$}
\end{center}

\setlength{\tabcolsep}{18.0pt}
\setlength{\extrarowheight}{1.0pt}
\begin{table}[th]
\begin{center}
\begin{tabular}{|c||c|c|c|}
\hline
\\[-23pt]
$B$ ($B_{{}_0}$)  & \multicolumn{3}{c|}{ $E$ } \\
\hline                  & \multicolumn{3}{l|}{\hspace{1.0cm} $s=0$ \hspace{3.2cm} $s=1$ \hspace{3.3cm} $s=2$ }\\
\hline
\hline
$\frac{1}{80}$                            &   $-1.99879415981(2)$    &  $-0.24362779(6)$      &   $-0.113195431(5)$
\\[3pt]
\hline
$\frac{1}{40}$                            &   $-1.9984430154(9)$     &  $-0.259036915(9)$     &   $-0.1280256740(7)$
\\[3pt]
\hline
$\frac{1}{22}$                            &   $-1.99736515(6)$       &  $-0.274591392(4)$     &   $-0.13667596638(9)$
\\[3pt]
\hline
$\frac{1}{10+\sqrt{73}}$                  &   $-1.996736659(3)$      &  $-0.27845077(5)$      &   $-0.13709579369(7)$
\\[3pt]
\hline
$\frac{1}{14}$                            &   $-1.99510269(8)$       &  $-0.282908684(7)$     &   $-0.13426953513(7)$
\\[3pt]
\hline
$\frac{1}{6}$                             &   $-1.97852822(1)$       &  $-0.259169175(9)$     &   $-0.0744506106(7)$
\\[3pt]
\hline
$\frac{1}{5}$                             &   $-1.96980803(9)$       &  $-0.239737216(2)$     &   $-0.0440478620(8)$
\\[3pt]
\hline
$\frac{1}{4}$                             &   $-1.9540953(6)$        &  $-0.204134835(5)$     &   $0.0067791680(5)$
\\[3pt]
\hline
$\frac{1}{3}$                             &   $-1.9214286(5)$        &  $-0.1324163386(5)$    &   $0.1013342701(0)$
\\[3pt]
\hline
$\frac{1}{10-\sqrt{73}}$                  &   $-1.711958(5)$         &   $0.2635196563(2)$    &   $0.573818518(4)$
\\[3pt]
\hline
$1$                                       &   $-1.4587925(9)$        &   $0.6770274379(6)$    &   $1.040528063(2)$
\\[3pt]
\hline
$\frac{5}{2}$                             &   $0.1847432(2)$         &   $2.952974280(0)$     &   $3.4998230(1)$
\\[3pt]
\hline
$\frac{25}{2}$                            &   $15.3697251(5)$        &  $20.51777694(3)$      &   $21.69669571(0)$
\\[3pt]
\hline
$25$                                      &   $36.7079218(0)$        &  $43.69741030(5)$      &   $45.3617446(1)$
\\[3pt]
\hline
$125$                                     &  $221.14240736(2)$       &  $236.0651014(3)$      &  $239.881963(3)$
\\[3pt]
\hline
$250$                                     &  $459.4105028(2)$        &  $480.4001931(5)$      &  $485.939216(8)$
\\[3pt]
\hline
$500  $                                   &  $942.73495448(3)$       & $972.4683323(0)$       &  $980.60230(1)$
\\[3pt]
\hline
\end{tabular}
\caption{\small Ground state energy $E$ in Hartrees (see (\ref{HKa})),
a modification due to the second correction $E_2$ (\ref{En0})
indicated by a number in brackets; magnetic field in effective atomic units,
$B_0 = 9.391766\times 10^9 \,G$. }
\label{Table6}
\end{center}
\end{table}

\clearpage

\begin{center}
\textbf{Hydrogen atom $(q=0)$ at $\mathbf P=0$. Total energy $E$}
\end{center}

\setlength{\tabcolsep}{16.0pt}
\setlength{\extrarowheight}{1.5pt}
\begin{table}[th]
\scriptsize
\begin{center}
\begin{tabular}{|c||c|c|c|}
\hline
\\[-16pt]
 $B$ ($B_{{}_0}$)  & \multicolumn{3}{c|}{ $E$  } \\
\hline                   & \multicolumn{3}{l|}{ \hspace{0.5cm} $s=0$ \hspace{2.4cm} $s=1$ \hspace{2.5cm} $s=2$   }\\
\hline
\hline
$\frac{1}{80}$                            &   $ -0.217635(9)  $  &  $ -0.0811667(6) $       & $ -0.0374540(5)   $
\\[3pt]
\hline
$\frac{1}{40}$                            &   $ -0.2051751(6) $  &  $ -0.0563387(9) $       & $ -0.00177977(5)  $
\\[3pt]
\hline
$\frac{1}{22}$                            &   $ -0.171909(7)  $  &  $  0.0057230(7) $       & $  0.073899775(0) $
\\[3pt]
\hline
$\frac{1}{10+\sqrt{73}}$                  &   $ -0.154594(6)  $  &  $  0.0356050(1) $       & $  0.108450073(3) $
\\[3pt]
\hline
$\frac{1}{14}$                            &   $ -0.113832(9)  $  &  $ 0.10207154(5) $       & $  0.183447316(5) $
\\[3pt]
\hline
$ \frac{1}{6}$                            &   $  0.180603(3)  $  &  $ 0.518754514(2) $      & $  0.633897911275(5)  $
\\[3pt]
\hline
$\frac{1}{5}$                             &   $  0.30169(0)   $  &  $ 0.676026877(6) $      & $  0.7995645073489^*  $
\\[3pt]
\hline
$\frac{1}{4}$                             &   $ 0.4944103(1)  $  &  $ 0.918148254(1) $      & $  1.05509437436(0)   $
\\[3pt]
\hline
$\frac{1}{3}$                             &   $   0.837351(0) $  &  $ 1.3326075122^* $      & $  1.4886452575(4)    $
\\[3pt]
\hline
$\frac{1}{10-\sqrt{73}}$                  &   $ 2.45501566(2) $  &  $ 3.178464838(5) $      & $  3.393569013(3)     $
\\[3pt]
\hline
$1$                                       &   $ 3.997822536^* $  &  $ 4.86961381(1)  $      & $  5.1255935499(3)    $
\\[3pt]
\hline
$\frac{5}{2}$                             &   $ 11.8771174(3) $  &  $ 13.231202656(3) $     & $  13.6268591463(3)   $
\\[3pt]
\hline
$\frac{25}{2}$                            &   $  68.152885(0) $  &  $ 71.078388655(1) $     & $  71.954727704(5)    $
\\[3pt]
\hline
$25$                                      &   $  140.356605(8)$  &  $ 144.459685385(0) $    & $  145.707155528(6)   $
\\[3pt]
\hline
$125$                                     &   $ 728.412926(6) $  &  $ 737.576787437(2) $    & $  740.49341090(5)    $
\\[3pt]
\hline
$250$                                     &   $ 1469.302175(1)$  &  $ 1482.36469154(6) $    & $  1486.6410268(2)    $
\\[3pt]
\hline
$500$                                     &   $ 2956.17917(8) $  &  $ 2974.91385507(8) $    & $  2981.2724746(3)    $
\\[3pt]
\hline
\end{tabular}
\caption{ First excited state energy $E$ in Hartrees (see (\ref{HKa})),
a modification due to the second correction $E_2$ (\ref{En0}) indicated by a number in brackets; magnetic field in effective atomic
units, $B_0 = 9.391766\times 10^9 \,G$. Energies corresponding to the exact solutions marked by * \,.}
\label{Table7}
\end{center}
\end{table}

\clearpage

\begin{center}
\textbf{Hydrogen atom $(q=0)$ at $\mathbf P=0$. Node for first excited state}
\end{center}

\setlength{\tabcolsep}{15.0pt}
\setlength{\extrarowheight}{2.5pt}
\begin{table}[th]
\begin{center}
\begin{tabular}{|c||c|c|c|}
\hline
\\[-24pt]
 $B$ ($B_{{}_0}$)  & \multicolumn{3}{c|}{$s=0$ \hspace{2.0cm} $s=1$ \hspace{2.1cm} $s=2$  } \\
\hline               & \multicolumn{3}{l|}{ \hspace{0.4cm} $f_0 +f_1$  \hspace{1.8cm}  $f_0 +f_1$ \hspace{1.7cm}  $f_0 +f_1$ }\\
\hline
\hline
$\frac{1}{80}$                            &   $ 0.7(4)$        &  $ 3.59(9)$          & $ 7.0209(8)$
\\[3pt]
\hline
$\frac{1}{40}$                            &   $ 0.748222(0)$   &  $ 3.345(1)$         & $ 5.770(3)$
\\[3pt]
\hline
$\frac{1}{22}$                            &   $ 0.7439(9)$     &  $ 2.9812(6)$        & $ 4.668(2)$
\\[3pt]
\hline
$\frac{1}{10+\sqrt{73}}$                  &   $ 0.741(8)$      &  $ 2.85769(4)$       & $ 4.3700(7)$
\\[3pt]
\hline
$\frac{1}{14}$                            &   $ 0.736(8)$      &  $ 2.64302(1)$       & $ 3.90417(1)$
\\[3pt]
\hline
$ \frac{1}{6}$                            &   $ 0.7041(9)$     &  $ 1.98125(0)$       & $ 2.71431(0)$
\\[3pt]
\hline
$\frac{1}{5}$                             &   $ 0.6922(5)$     &  $ 1.84704(2)$       & $ 2.501361655^*$
\\[3pt]
\hline
$\frac{1}{4}$                             &   $ 0.674(7)$      &  $ 1.690022(6)$      & $ 2.2604603(0)$
\\[3pt]
\hline
$\frac{1}{3}$                             &   $ 0.6471(1)$     &  $ 1.500816993^* $   & $ 1.9802548(2)$
\\[3pt]
\hline
$\frac{1}{10-\sqrt{73}}$                  &   $ 0.55581(3)$    &  $ 1.09590(5)$       & $ 1.410159(8)$
\\[3pt]
\hline
$1$                                       &   $ 0.500270665^*$ &  $ 0.92442(8)$       & $ 1.178530(8)$
\\[3pt]
\hline
$\frac{5}{2}$                             &   $ 0.36421(3) $   &  $ 0.6026555(7)$     & $ 0.756376(0)$
\\[3pt]
\hline
$\frac{25}{2}$                            &   $ 0.18375(4)$    &  $ 0.277029(6)$      & $ 0.342891(7)$
\\[3pt]
\hline
$25$                                      &   $ 0.13337(8)$    &  $ 0.19713(2)$       & $ 0.243231(3)$
\\[3pt]
\hline
$125$                                     &   $ 0.061673(2)$   &  $ 0.088898(2)$      & $ 0.109234(6)$
\\[3pt]
\hline
$250$                                     &   $ 0.0439445(7)$  &  $ 0.0629836(7)$     & $ 0.077317(1)$
\\[3pt]
\hline
$500$                                     &   $ 0.0312402(2)$  &  $ 0.0445976(4) $    & $ 0.054709(7)$
\\[3pt]
\hline
\end{tabular}
\caption{\small First excited state: evolution of node $f$ in a.u.,
modification due to the first correction $f_1$ (\ref{PTn}) indicated by a number in brackets.
$B_0= 9.391766\times 10^9 \,G$.}
\label{Table8}
\end{center}
\end{table}

\clearpage

\begin{center}
\textbf{Hydrogen atom $(q=0)$ at $\mathbf P=0$. Total energy $E$}
\end{center}

\setlength{\tabcolsep}{15.0pt}
\setlength{\extrarowheight}{2.5pt}
\begin{table}[th]
\scriptsize
\begin{center}
\begin{tabular}{|c||c|c|c|}
\hline
\\[-17pt]
 $B$ ($B_{{}_0}$)  & \multicolumn{3}{c|}{ $E$  } \\
\hline                   & \multicolumn{3}{l|}{\hspace{0.5cm} $s=0$ \hspace{2.5cm} $s=1$ \hspace{2.7cm} $s=2$  }\\
\hline
\hline
$\frac{1}{80}$                            &   $ -0.053761(8)  $   &  $ -0.00267077(7) $     &  $ 0.024860663(1)   $
\\[3pt]
\hline
$\frac{1}{40}$                            &   $  0.001633(7)  $   &  $ 0.07487168(0)  $     &  $ 0.1125652633(1)  $
\\[3pt]
\hline
$\frac{1}{22}$                            &   $  0.1176832(0) $   &  $ 0.22338150793(6)$    &  $ 0.2727766527706^*$
\\[3pt]
\hline
$\frac{1}{10+\sqrt{73}}$                  &   $  0.17099053(2) $  &  $ 0.288844238(8) $     &  $ 0.3421829713(5)  $
\\[3pt]
\hline
$\frac{1}{14}$                            &   $  0.2874008(8)  $  &  $ 0.428493577359699^*$ &  $ 0.489006320917(3) $
\\[3pt]
\hline
$ \frac{1}{6}$                            &   $  0.9994556331^* $ &  $ 1.2391284239(5)  $   &  $ 1.3278640075(8)   $
\\[3pt]
\hline
$\frac{1}{5}$                             &   $  1.2661322(1) $   &  $ 1.5333587264(7)  $   &  $ 1.62980988910(9)  $
\\[3pt]
\hline
$\frac{1}{4}$                             &   $  1.67605782(6) $  &  $ 1.9804015571(6)  $   &  $ 2.087281521(1)    $
\\[3pt]
\hline
$\frac{1}{3}$                             &   $  2.37806227(8) $  &  $ 2.7361449127(8)$     &  $ 2.8582812169(3)   $
\\[3pt]
\hline
$\frac{1}{10-\sqrt{73}}$                  &   $  5.491528597^* $  &  $ 6.01868884(1)  $     &  $ 6.1903994984(6)   $
\\[3pt]
\hline
$1$                                       &   $  8.33980704(9) $  &  $ 8.978510103(0) $     &  $ 9.1840767428(4)   $
\\[3pt]
\hline
$\frac{5}{2} $                            &   $ 22.38619980(2) $  &  $ 23.39343405(5)   $   &  $ 23.714702525(7)   $
\\[3pt]
\hline
$\frac{25}{2}$                            &   $ 119.183572(1) $   &  $ 121.409837831(9) $   &  $ 122.13070108(7)   $
\\[3pt]
\hline
$25$                                      &   $ 241.762627(0) $   &  $ 244.90716668(6)  $   &  $ 245.93830041(8)   $
\\[3pt]
\hline
$125$                                     &   $ 1231.296005(7) $  &  $ 1238.4112545(8)  $   &  $ 1240.854558(7)    $
\\[3pt]
\hline
$250  $                                   &   $ 2473.182285(5) $  &  $ 2483.3801579(9)  $   &  $ 2486.9905782(9)   $
\\[3pt]
\hline
$500$                                     &   $ 4961.309675(1) $  &  $ 4976.0258668(4)  $   &  $ 4981.446188(9)    $
\\[3pt]
\hline
\end{tabular}
\caption{\small The second excited state energy $E$ in Hartrees(see (\ref{HKa})),
a modification due to the second correction $E_2$ (\ref{En0}) indicated by a number in brackets;
magnetic field in effective atomic units, $B_0 = 9.391766\times 10^9 \,G$. Energies corresponding to
the exact solutions marked by * .}
\label{Table9}
\end{center}
\end{table}

\clearpage

\begin{center}
\textbf{Hydrogen atom $(q=0)$ at $\mathbf P=0$. Nodes $f^{a(b)}$ (a.u.) for 2nd-excited state}
\end{center}

\setlength{\tabcolsep}{4.0pt}
\setlength{\extrarowheight}{2.5pt}
\begin{table}[th]
\scriptsize
\begin{center}
\begin{tabular}{|c||c|c|c|c|c|c|}
\hline
 \\[-17pt]
 $B$ ($B_{{}_0}$)  & \multicolumn{6}{c|}{      $s=0$   \hspace{4.5cm}     $s=1$        \hspace{4.5cm}     $s=2$   } \\[2pt]
\hline  & \multicolumn{6}{c|}{ \hspace{0.8cm} $f^a_0 +f^a_{1}$  \hspace{1.6cm}  $f^b_0 +f^b_{1}$ \hspace{1.6cm}    $f^a_0 +f^a_{1}$  \hspace{1.5cm}  $f^b_0 +f^b_{1}$ \hspace{1.5cm}  $f^a_0 +f^a_{1}$  \hspace{1.4cm}  $f^b_0 +f^b_{1}$\hspace{0.8cm} }\\[2pt]
\hline
\hline
$\frac{1}{80}$                    &   $ 0.7(3)              $  &   $ 4.1(1) $              &$ 3.(1)$              &$ 8.4(8)  $            &$ 5.(7) $             &$  11.(9)  $
\\[2pt]
\hline
$\frac{1}{40}$                    &   $ 0.723(2)            $  &   $ 3.8(2)  $             &$ 2.87(8) $           &$ 7.0(3)  $            &$ 4.7(9)  $           &$ 9.1(6)   $
\\[2pt]
\hline
$\frac{1}{22}$                    &   $ 0.7103(0)           $  &   $ 3.441(0) $            &$ 2.496(0) $          &$ 5.7(3)  $            &$ 3.8437800^*  $      &$ 7.162211258^*  $
\\[2pt]
\hline
$\frac{1}{10+\sqrt{73}}$          &   $ 0.704(3)            $  &   $ 3.309(3) $            &$ 2.377(4) $          &$ 5.38(2) $            &$ 3.593(2) $          &$ 6.65(0)  $
\\[2pt]
\hline
$\frac{1}{14}$                    &   $ 0.6930(6)           $  &   $ 3.080(3)  $           &$ 2.17831014^* $      &$ 4.82550249^* $       &$ 3.204(1) $          &$ 5.87(5)  $
\\[2pt]
\hline
$ \frac{1}{6}$                    &   $ 0.63431989828^*     $  &   $ 2.3673140886^*  $     &$ 1.6039(5) $         &$ 3.377(0) $           &$ 2.220(6) $          &$ 3.988(6) $
\\[2pt]
\hline
$\frac{1}{5}$                     &   $ 0.6164(9)           $  &   $ 2.2(1) $              &$ 1.491(7) $          &$ 3.1153(3)$           &$ 2.045(6) $          &$ 3.662(1) $
\\[2pt]
\hline
$\frac{1}{4}$                     &   $ 0.5922(3)           $  &   $ 2.0460(3) $           &$ 1.3615(8) $         &$ 2.818(3) $           &$ 1.8480(3) $         &$ 3.2961(5)  $
\\[2pt]
\hline
$\frac{1}{3}$                     &   $ 0.5574(6)           $  &   $ 1.8335(6) $           &$ 1.20614(6) $        &$ 2.47198(8)  $        &$ 1.618(3) $          &$ 2.87472(3) $
\\[2pt]
\hline
$\frac{1}{10-\sqrt{73}}$          &   $ 0.4589805381^*      $  &   $ 1.36608961752^*      $&$ 0.877202(9) $       &$ 1.76412(8)  $        &$ 1.151(8) $          &$ 2.02998(3) $
\\[2pt]
\hline
$1$                               &   $ 0.40609(6)          $  &   $ 1.162117(2) $         &$ 0.739001(2) $       &$ 1.475503(9) $        &$ 0.9625(4) $         &$ 1.69113(6) $
\\[2pt]
\hline
$\frac{5}{2} $                    &   $ 0.287531(2)         $  &   $ 0.769136(5) $         &$ 0.480869(0) $       &$ 0.948181(7) $        &$ 0.617646(0) $       &$ 1.079389(9)    $
\\[2pt]
\hline
$\frac{25}{2}$                    &   $ 0.14216347(3)       $  &   $ 0.35840020(2)  $      &$ 0.22075161(6) $     &$ 0.43031012(5) $      &$ 0.2799747(5) $      &$ 0.48685198(7)  $
\\[2pt]
\hline
$25  $                            &   $ 0.102810770(2)      $  &   $ 0.255806360(3)  $     &$ 0.157045454(6) $    &$ 0.30531454(7) $      &$ 0.19859894(7) $     &$ 0.344944567(6) $
\\[2pt]
\hline
$125   $                          &   $ 0.0473384940(6)     $  &   $ 0.11580464450(5) $    &$ 0.070798731(0)$     &$ 0.13715803759(9)  $  &$ 0.0891896671(3) $   &$ 0.1546734540(4)   $
\\[2pt]
\hline
$250  $                           &   $ 0.033700437982(7)   $  &   $ 0.082119553759(6) $   &$ 0.0501569017(7) $   &$  0.09708842652(6) $  &$ 0.06312912347(2) $  &$ 0.10943912950(2)  $
\\[2pt]
\hline
$500  $                           &   $ 0.02394317182217(9) $  &   $ 0.0581835897196(9) $  &$ 0.035513575575(7) $ &$ 0.068703348311(1) $  &$ 0.04467026830(5) $  &$ 0.077419369782(5) $
\\[2pt]
\hline
\end{tabular}
\caption{Modification of the node position due to the correction $f_1$ (\ref{PTn}) indicated
by a number in brackets.}
\label{Table10}
\end{center}
\end{table}


\begin{table}[htb]
\begin{center}
\begin{tabular}{|c||c|c|}
\hline
$B\, (a.u.)\,$       &   $E$      &  $E^{AIM}$  \\ \hline
\hline
$\frac{1}{10}$       &  $ -1.999531$        &  $-1.999530$        \\
\hline
$\frac{1}{4}$        &  $ -1.997079$        &  $-1.997078$        \\
\hline
$\frac{107}{250}$    &  $ -1.991490$        &  $-1.991490$        \\
\hline
$1$                  &  $ -1.955159$        &  $-1.955159$        \\
\hline
\hline
\end{tabular}
\end{center}
\caption{Hydrogen atom with infinitely heavy proton: The ground-state energy (in Hartrees) from the present study $E$ in comparison with $E^{AIM}$ obtained by A. Soylu  \textit{et al}. \cite{Soy}, $B_0=2.3505 \times 10^9 G$. }
\label{qComparison}
\end{table}

\end{document}